\documentclass[12pt]{article}
\usepackage[utf8]{inputenc}

\usepackage{graphicx} 
\usepackage{amsmath,amssymb,amsfonts}
\usepackage{xcolor}
\definecolor{fqxicolor}{HTML}{bf9000}

\usepackage[bookmarks=true,
    pdfnewwindow=true,      
    colorlinks=true,        
    linkcolor=fqxicolor,    
    citecolor=fqxicolor,    
    filecolor=fqxicolor,    
    urlcolor=fqxicolor,     
    final=true
]{hyperref}

\usepackage[manuscript,sequential-numbered]{fqxi_foundational_review}
\setcitestyle{aysep={}}

\usepackage[utf8]{inputenc}
\usepackage[T1]{fontenc}
\usepackage{qcircuit}
\usepackage{comment}
\usepackage{float} 
\usepackage{pdflscape}

\usepackage{booktabs}
\usepackage{tabularx}
\usepackage{caption}
\usepackage{array}
\newcolumntype{L}[1]{>{\raggedright\arraybackslash}p{#1}}
\newcolumntype{Y}{>{\raggedright\arraybackslash}X}

\usepackage{orcidlink}

\newcommand{\abs}[1]{\left| #1 \right|}
\renewcommand{\Re}{\mathfrak{R}}
\renewcommand{\Im}{\mathfrak{I}}
\newcommand{\bra}[1]{\left\langle #1\right|}
\newcommand{\ave}[1]{\langle #1\rangle}  
\newcommand{\ket}[1]{\left|#1\right\rangle}
\newcommand{\ketbra}[2]{\left|#1\right\rangle\!\left\langle #2\right|}
\newcommand{\braket}[2]{\left\langle #1|#2\right\rangle}
\newcommand{\bramatket}[3]{\langle #1 \hspace{1pt} | #2 | \hspace{1pt} #3 \rangle}
\newcommand{\ketbrad}[1]{|#1\rangle\!\langle #1|}
\newcommand{\tr}[1]{\mathrm{tr}\left\{#1\right\}}
\newcommand{\ptr}[2]{\mathrm{tr_{#1}}\left\{#2\right\}}

\newcommand{\la}{\left\langle}
\newcommand{\ra}{\right\rangle}
\newcommand{\pd}{\partial}
\newcommand{\de}[1]{\delta\left(#1\right)}

\newcommand{\e}[1]{\exp{\left(#1\right)}}

\newcommand{\lo}[1]{\ln{\left(#1\right)}}
\newcommand{\id}{\mathbb{I}}
\newcommand{\com}[2]{\left[#1,\,#2\right]}

\newcommand{\bla}{bla\\bla\\bla\\bla\\bla}

\newcommand{\mc}[1]{\mathcal{#1}}

\newcommand{\mrm}[1]{\mathrm{#1}}

\newcommand{\dbar}{d\hspace*{-0.08em}\bar{}\hspace*{0.1em}}

\usepackage[none]{hyphenat}

\title{Thermodynamic work in the quantum regime}

\author[1,2]{Tara Kalsi \orcidlink{0000-0003-1513-5795} \thanks{\href{mail to:tkalsi@umbc.edu}{tkalsi@umbc.edu}}}
\author[3,4]{Akira Sone \orcidlink{0000-0003-3539-6140} \thanks{\href{mail to:Akira.Sone@umb.edu}{akira.sone@umb.edu}}}
\author[5]{Marcus V. S. Bonan\c{c}a \orcidlink{0000-0001-7283-3034} \thanks{\href{mail to:mbonanca@unicamp.br}{mbonanca@unicamp.br}}}
\author[1,2,6]{Sebastian Deffner \orcidlink{0000-0003-0504-6932} \thanks{\href{mail to:deffner@umbc.edu}{deffner@umbc.edu}} }

\affil[1]{Department of Physics, University of Maryland, Baltimore County, Baltimore, Maryland 21250, USA}
\affil[2]{Quantum Science Institute, University of Maryland, Baltimore County, Baltimore, Maryland 21250, USA}
\affil[3]{Department of Physics, University of Massachusetts, Boston, Massachusetts 02125, USA}
\affil[4]{The NSF AI Institute for Artificial Intelligence and Fundamental Interactions}
\affil[5]{Gleb Wataghin Physics Institute, The University of Campinas - 13083-859, Campinas, São Paulo, Brazil}
\affil[6]{National Quantum Laboratory, College Park, Maryland 20740, USA}
\date{}

\begin{document}

\maketitle

\begin{Abstract}
In the English language, there is a myriad of synonyms for the word \emph{work}, such as labor, task, job, chore, onus, etc. Yet, in theories of classical physics, work is not only one of the most central, but also most unambiguously defined concepts that is often expressed as \emph{force times displacement}. Curiously, the quantum domain appears to be more akin to the English language than to classical physics in the sense that a plethora of different notions of \emph{quantum work} have been proposed. This review summarizes and discusses the most important notions that have had a significant impact in the literature. Particular emphasis is put on pedagogical accessibility and a mostly self-consistent presentation of one of the most fundamental topics in modern quantum physics.
\end{Abstract}

\newpage

\tableofcontents

\newpage
\section{Introduction}

The concept of \emph{work} is intrinsically linked to the nascence of physics as a discipline of study. Its first precise definition can be traced back to Coriolis~\cite{Coriolis1829}, who introduced the now-common notion of ``force times displacement.'' Especially in the theory of classical mechanics, work is the central quantity upon which all other concepts are defined. For instance, \emph{energy} is defined as the work stored in a physical system, and \emph{momentum} can be expressed in terms of work via the energy-momentum relation.

In thermodynamics, it is then recognized that \emph{mechanical work} is only one part of the contributions to changes in the internal energy~\cite{Callen1985}. In fact, work is understood as the amount of \emph{internal energy} that can be exchanged in a known and controlled manner, whereas \emph{heat} refers to the uncontrolled exchange of energy with the rest of the universe. 

While thermodynamics is a phenomenological theory that describes the average behavior of macroscopic observables, classical mechanics~\cite{Goldstein2011} is a microscopic theory of trajectories and dynamics. Hence, these two theories are typically understood to describe very different physical scenarios: 
thermodynamics is concerned with the state transformations 
and exchange of energy between macroscopic systems, whereas classical mechanics is interested in the microscopic dynamics of individual or collections of particles. The two theories meet in the study of thermodynamic properties of small systems, whose dynamics are governed by fluctuations. Importantly, no matter whether the dynamics is deterministic or stochastic, the notion of work remains the same, namely ``force times displacement''~\cite{Jarzynski1997PRL,Jarzynski1997PRE,Sekimoto1998,Chernyak2006JSM}. However, especially in stochastic dynamics, it is instrumental to be precise about how the dynamics is described. For instance, it is a well-known fact that depending on how the Langevin equation is discretized, different Fokker-Planck equations can be obtained~\cite{risken1989fokker}. 

The situation becomes even more involved for quantum systems. Quantum systems can also be described in terms of thermodynamics, yet classical mechanics has to be replaced by quantum mechanics. Consequently, the Hamiltonian equations of motion are replaced by the Schr\"odinger equation, and hence the concept of a ``classical trajectory'' becomes somewhat vacuous. If, however, ``displacement'' is no longer well-defined, how do we then define \emph{quantum work}? And if we do not know how to define work, how do we construct a sound theory of \emph{quantum thermodynamics}~\cite{Deffner2019book}? 

The present review outlines the major ideas, concepts, and paradigms that have been proposed as definitions of \emph{quantum work}~\cite{chua2026looking}. As we will see, over the last 25 years several rather distinct notions have been considered, from conceptually simple and experimentally relevant to mathematically involved yet physically and universally valid for a large range of scenarios~\cite{Silva2021PRA,Yukawa2000JPSJ,Chernyak2004PRL,Allahverdyan2005PRE,Engel2007EPL,Hanggi2015NP,Liu2014CPB,Sone2025,Vijay2026NJP,Xu2020PRA,Cohen2012PRE,Mohammady2021PRA,Silva2023PRA}. The major complications rest in the genuine non-classical nature of quantum fluctuations. While it is easy to see that in classical dynamics pretty much any definition of work will give rise to the same statistics~\cite{Talkner2016PRE}, this is not true in the quantum domain. Rather, each and every notion of quantum work gives rise to unique statistics, some of which have measurable consequences.

In the following, we will outline the major formulations of quantum work that have found resonance in the literature. To this end, we will briefly summarize how thermodynamic work is defined in fluctuating, yet classical systems, before we present a somewhat comprehensive summary of quantum notions. The exposition is rounded off with a discussion of experimental demonstrations and realizations of the various approaches. While such \emph{foundational reviews} always strive for completeness, in reality authors inevitably have to make some decisions on what to cover and what to omit. Thus, we consider the present review only as a springboard for a deeper dive into a fascinating topic at the core of modern physics.

\section{Thermodynamic work in classical systems}

The concept of \emph{state} as the minimal amount of information necessary to determine all the relevant instantaneous properties of a given physical system is an instrumental concept at the heart of modern physics. In particular, the so-called \emph{thermodynamic} state of a macroscopic physical system (comprised of some macroscopic amount of matter) is the one that admits a classical thermodynamic description. The particular classical framework used depends on the choice of coarse-graining. 

Phenomenological thermodynamics is the broad framework in which we neglect the specific details of the individual atoms or particles comprising the system, and focus only on the equilibrium macroscopic properties. 
By contrast, stochastic thermodynamics bridges the emergence of the bulk, macroscopic thermodynamic properties of a state with the microscopic fluctuations of the individual constituents of the system, and extends into the non-equilibrium regime. 
Importantly, the average results are the same; only the level of detail with which we describe the system differs. 
 
\subsection{Phenomenological thermodynamics}

Phenomenological thermodynamics builds upon general principles and experimentally observed facts (such as the laws of thermodynamics), and does not assume or rely on microscopic (atomic or molecular) models; it applies universally.

Since phenomenological thermodynamics describes macroscopic, observable properties (i.e., ensemble averages over several microscopic states) of a system in thermal equilibrium, parameters that characterize the system of interest are state functions (pressure $P$, volume $V$, temperature $T$, internal energy $E$, thermodynamic entropy $S$). 
\textit{Intensive} variables are constant with respect to the amount of matter (e.g., density, pressure, temperature), while \textit{extensive} variables scale with system size (e.g., mass, total internal energy, volume) and are additive for subsystems.

\textit{Closed} systems can exchange energy (through \textit{heat} or \textit{work}) with their surroundings, but cannot exchange matter. 
If a closed system does not exchange energy with its surroundings, it is \textit{isolated}.
Conversely, \textit{open} systems can exchange both energy and matter with their surroundings.

For a system to be in \textit{thermodynamic equilibrium}, it must simultaneously be in thermal, mechanical, chemical, and radiative equilibrium~\cite{Callen1985}, so that there is no net macroscopic transfer of energy or matter. 
Macroscopic properties are therefore constant over time, so that the system is characterized by a spatially uniform temperature. 
Thermodynamic equilibrium will persist indefinitely, unless the system is disturbed by some thermodynamic operation. 

An \textit{ideal} thermodynamic process is free of dissipation. A \textit{quasistatic} process is an ideal thermodynamic process in which the system is brought from its initial state $A$ to the final state $B$ infinitely slowly, such that all intermediate states in the process are in thermodynamic equilibrium and intensive variables are uniform and well-defined. 
\textit{Adiabatic} (no heat transfer, $\dbar Q=0$), \textit{isothermal} (constant temperature, $dT=0$), \textit{isobaric} (constant pressure, $dP=0$), and \textit{isochoric} or \textit{isometric} (constant volume, $dV=0$) processes are commonly idealized as being quasistatic. 
The canonical example of a non-quasistatic process is the free expansion of a gas into a vacuum, during which pressure and temperature are ill-defined.   

A \textit{reversible} process is necessarily a quasistatic process, whose direction may be reversed by infinitesimal changes in the properties of the surroundings. 
Consequently, reversible processes do not take place in isolated systems. 
A quasistatic adiabatic process with no dissipative effects is reversible, and therefore an \textit{isentropic} process (characterized by constant \textit{thermodynamic entropy}, $dS=0$, defined below). 
Not all quasistatic processes are reversible, since the process can move slowly enough to remain in equilibrium, but still involve dissipative forces or losses (such as friction or heat flow across a finite temperature difference). 

Any \emph{equilibrium} process is then idealized as reversible, where changes are defined entirely by initial and final states, independent of the specific path taken.

\paragraph{The zeroth law and thermal equilibrium.}
The zeroth law of thermodynamics defines a state of equilibrium of a system relative to its environment. It is most commonly formulated in terms of three systems $A$, $B$, and $C$. If systems $A$ and $B$ are each independently in thermal equilibrium with system $C$, then systems $A$ and $B$ are in thermal equilibrium with each other.

Such equilibrium states are fully characterized by equations of state, which relate three experimentally accessible parameters, only two of which are independent. 
For conventional systems such as normal fluids, these parameters are typically pressure $P$, temperature $T$, and volume $V$, and their functional relationship depends on the specifics of the system, 
\begin{equation} \label{eq:equation-of-state}
    f(P,T,V) = 0. 
\end{equation}

Except at phase transitions, the function $f$ is taken to be continuously differentiable, so that the equation of state can be represented as a smooth surface (a thermodynamic manifold) in a three-dimensional space. 
Since quasistatic processes are a sequence of equilibrium states, they correspond to paths on the surface spanned by the equation of state. 
A process whose path contains a segment of nonzero length that does not lie on this manifold is necessarily irreversible. 
The converse does not hold, in the sense that quasistatic processes may still be irreversible if they involve dissipation. 
By definition, reversible processes are dissipation-free quasistatic processes, which are therefore represented by paths that lie entirely in the thermodynamic manifold, and can be traversed in either direction. 

\paragraph{The first law and work.}
The first law is a statement of energy conservation.  For an arbitrary thermodynamic process, the infinitesimal change in the internal energy $dE$ is given by 
\begin{equation} \label{eq:first-law}
    dE = \dbar W + \dbar Q, 
\end{equation}
where $\dbar W$ and $\dbar Q$ are also forms of infinitesimal energy changes, namely work $W$ and heat $Q$. 
Work $W$ represents the macroscopic energy transfer as a result of changes in extensive variables, and heat $Q$ represents the energy transferred due to the existence of a temperature gradient. 
These are dependent on the particular thermodynamic process, and therefore expressed in terms of inexact differentials. 
This is in contrast with the exact differential $dE$, which does not depend on the path traced in the thermodynamic manifold. 
Technically, we say that $E$ is a \textit{state function}, while $W$ and $Q$ are not. 

Convention typically defines work done \textit{by} the system on its surroundings to be positive, and work done \textit{on} the system by its surroundings to be negative. 
The typical example of a work protocol is that done on or by some gas contained in a cylinder with a movable piston, more precisely, a closed hydrostatic system undergoing some quasistatic, frictionless process, on which the infinitesimal work done is $\dbar W=-P\,dV$. 
Generalized work $\dbar W = -X\,dY$ is expressed in terms of a generalized force $X$ (typically an intensive parameter, like pressure or the magnetic field) and a generalized displacement $Y$ (an extensive parameter like volume or magnetization). 
Since the total work done on or by the system depends on an extensive variable, work scales with system size. 
For a closed system, the work done on the system by an external force is given by the difference between the energy at the initial and final (internal) points of the protocol. 

\paragraph{The second law and entropy.}
The second law further specifies the concept of heat $Q$ by introducing the notion of entropy. For a (closed) system in contact with a thermal bath at temperature $T$, an infinitesimal change in the thermodynamic entropy $dS$ satisfies
\begin{align} \label{eq:second-law}
    dS &\geq \frac{\dbar Q}{T}, 
\end{align}
where the inequality holds only if the process is reversible. 
If the process is irreversible, the entropy change (entropy production) 
$dS - \,\dbar Q/T$ is strictly positive.

In the limit that the system is isolated, no heat is exchanged with its surroundings, and $dS \geq 0$.  
Since entropy is a state function, its value only depends on the equilibrium state of the system and is independent of the process by which the state was attained, so the condition $dS\geq0$ is therefore a constraint on which states an isolated system can evolve into. 
In classical thermodynamics, entropy is a property of a thermodynamic system that expresses the direction or outcome of \textit{spontaneous} changes in the system: any \textit{spontaneous} process must either increase the entropy or leave it unchanged. 
The stationary equilibrium state of an isolated system maximizes entropy under its constraints. 

This asymmetry is typically referred to as the thermodynamic \textit{arrow of time}. 
Boltzmann's relation $S=k_B \ln{\mathcal{N}}$, where $\mathcal{N}$ is the number of microstates associated with a given macrostate and $k_B$ is the Boltzmann constant that we henceforth set to 1, enables a statistical interpretation, in which high-energy macrostates correspond to more microstates when compared to their low-entropy counterparts, so a system evolving in its phase space is significantly more likely to move towards higher entropy. 
Importantly, the microscopic dynamics are symmetric under time-reversal symmetry, so this irreversibility is statistical rather than dynamical.

Integrating the differential form~\eqref{eq:second-law} along a process connecting two equilibrium states $A$ and $B$ on the thermodynamic manifold, yields the \textit{Clausius inequality}
\begin{equation}
    \Delta S_{A\to B} \geq \int_A^B \frac{\dbar Q}{T}, 
\end{equation}
which bounds the entropy change between $A$ and $B$. 
The integral is performed along the process, and $T$ is the temperature of the thermal bath with which the system exchanges heat. More generally, the second law is expressed through several equivalent statements (e.g., Kelvin-Planck~\cite{PlanckBook} and Clausius~\cite{Clausius}, in addition to Carnot's theorem on the maximal efficiency of heat engines~\cite{Callen1985}).

Having defined the entropy, one can introduce the remaining \textit{thermodynamic potentials} via Legendre transformations. 
The Helmholtz free energy is defined via
\begin{equation} \label{eq:helmholtz-free-energy}
    F=E-TS, 
\end{equation}
while the Gibbs free energy is given by 
\begin{equation} \label{eq:gibbs-free-energy}
    G=H-TS=E+PV-TS. 
\end{equation}
The integral of the first law~\eqref{eq:first-law} is the internal energy $E=Q+W$, which, along with the enthalpy $H=E+PV$, in addition to the Helmholtz and Gibbs free energies, constitutes the four primary thermodynamic potentials for a simple single-phase system in standard classical thermodynamics. 

\paragraph{The third law and zero temperature.}
The third law (also known as the Nernst theorem~\cite{Callen1985}) describes how the entropy of a closed system at thermodynamic equilibrium is independent of parameters characterizing the system and therefore approaches a constant value when its temperature approaches absolute zero. 
At absolute zero (0K), the state of the system is that which minimizes the energy.
More precisely, the Nernst theorem states that as the temperature of a system approaches absolute zero, the entropy change for a physical transformation vanishes, 
\begin{equation} 
    \lim_{T\to0} \Delta S = 0. 
\end{equation}

\paragraph{The Onsager relations as the fourth law.}
Up until now, the first few laws have been formulated for an \textit{equilibrium} theory of thermodynamics. 
The so-called fourth law begins to look towards a \textit{non-equilibrium} theory that describes how heat and matter flow when a system is near, but not completely at, equilibrium, provided that thermodynamic variables can be defined locally under the assumption of \textit{local equilibrium}. 

Since physical systems are not typically in isotropic and homogeneous states of equilibrium, thermodynamic properties typically vary in space and time. 
Provided that these variations are weak over macroscopic scales, the system may be thought of as being in a state of \textit{local} equilibrium, so that for any point in space and time, thermodynamic quantities like the local (particle) density $n(\vec{r},t)$ and local energy density $e(\vec{r},t)$, together with intensive variables like the local temperature $T(\vec{r},t)$, can be defined.

It is instructive to consider the dynamics of the entropy $S=S(t)$, whose time derivative can be expressed in terms of the extensive quantities $\{X_k\}$ (e.g., energy, number of particles) that vary as a function of time, 
\begin{equation}
    \frac{dS}{dt} = \sum_k \frac{\partial S}{\partial X_k} \frac{dX_k}{dt} = \sum_k  \mathsf{F}_k \frac{dX_k}{dt}, 
\end{equation}
where $\mathsf{F}_k \equiv \partial S/\partial X_k$ are the entropy-representation intensive parameters (e.g., $1/T$ for energy $E$). 
For an isolated composite system of two subsystems exchanging $X_k$ across a boundary, conservation gives $dX_k^{(1)} = -dX_k^{(2)}$, so that 
\begin{equation}
    \frac{dS}{dt} = \sum_k  \mathcal{F}_k J_k \geq 0, 
\end{equation}
where the rates $J_k \equiv dX_k^{(2)}/dt$ define the thermodynamic \textit{fluxes}, and the differences $\mathcal{F}_k \equiv \mathsf{F}_k^{(2)} - \mathsf{F}_k^{(1)}$ define the thermodynamic \textit{forces} or \textit{affinities}.\footnote{Following Ref.~\cite{Callen1985}, an affinity is a generalized force, which, for \textit{discrete} systems, is expressed as the difference in the entropy-representation intensive parameters. For \textit{continuous} systems, it is the gradient of the entropy-representation intensive parameters.}
The rate of entropy production is the sum of products of each flux with its associated affinity. 

Alternatively, we can consider the dynamics of the local entropy $s(\vec{r},t)$, whose time derivative can be expressed in terms of the local densities $\{x_k\}$ of the extensive quantities that vary as a function of time, 
\begin{equation}
    \frac{\partial s}{\partial t} = \sum_k \frac{\partial s}{\partial x_k} \frac{ \partial x_k}{\partial t} = \sum_k \mathsf{F}_k \frac{ \partial x_k}{\partial t}, 
\end{equation}
where $\mathsf{F}_k \equiv \partial s/\partial x_k = \partial S/\partial X_k$ are the same intensive parameters as above. 
For conserved quantities, each density obeys a continuity equation with an associated thermodynamic flux $\vec{J}_k$
\begin{equation}
  \frac{{\partial}x_k}{{\partial} t} = - \nabla\cdot \vec{J}_k. 
\end{equation}
Substituting and using the product rule, we find 
\begin{equation}
    \frac{\partial s}{\partial t} = -\nabla \cdot \left( \sum_k \mathsf{F}_k \vec{J}_k\right) + \sum_k \vec{J}_k \cdot \nabla \mathsf{F}_k, 
\end{equation}
where the first term is the divergence of the entropy flux $\vec{J}_s = \sum_k \mathsf{F}_k \vec{J}_k$, characterizing the entropy flowing in or out of a region, and the second term is the local entropy production 
\begin{equation}
\label{eq:sigma}
    {\sigma} = \sum_k \vec{J}_k \cdot \nabla \mathsf{F}_k \geq 0, 
\end{equation}
which the second law requires to be nonnegative. 
Here, the gradients $\nabla \mathsf{F}_k$ act as the affinities driving the conjugate fluxes $\vec{J}_k$. 
In general, the fluxes are complicated functions. 
A simplifying case is purely resistive systems, for which the definition of the local flux only depends on the instantaneous local affinities. 
For small affinities, i.e., close to equilibrium, the fluxes can be expanded in powers of the affinities. 
In leading order, 
\begin{align}
    J_k&=\sum_j L_{jk} \mathcal{F}_j, 
\end{align}
where the kinetic coefficients 
\begin{equation}
    L_{jk} \equiv \left.\frac{\partial J_k}{\partial \mathcal{F}_j}\right|_{\mathcal{F}_j=0}
\end{equation}
are evaluated at equilibrium, so that all affinities are vanishing. 

The Onsager relation~\cite{Onsager1931ReciprocalI, Onsager1931ReciprocalII} states, under microscopic reversibility, i.e., time-reversibility in microscopic dynamics (e.g., in the absence of external magnetic fields), that the matrix of kinetic coefficients is symmetric, 
\begin{equation}
    L_{jk} = L_{kj}, 
\end{equation}
so that the coefficient coupling the force (affinity) $j$ on flux $k$ is equal to the coefficient coupling the force (affinity) $k$ on flux $j$. 
To understand what this means physically, consider the dynamics of a fluid system, for which temperature differences drive the flow of heat from warmer regions to cooler regions. 
Likewise, pressure differences drive the flow of matter from high-pressure regions to low-pressure regions. 
When both temperature and pressure vary in a system, the Onsager relation states that pressure differences at constant temperature can drive the flow of heat, and likewise, temperature differences at constant pressure can drive the flow of matter. 
In this reciprocal way, the Onsager relation can be interpreted as the thermodynamic analog of  Newton's third law.

When time-reversal symmetry is broken at the microscopic level, e.g., by an external magnetic field $\mathbf{B}$, the relations are generalized to the Onsager-Casimir relations~\cite{Casimir1945}, 
\begin{equation}
    L_{jk}(\mathbf{B}) = \epsilon_j \epsilon_k L_{kj}(-\mathbf{B}), 
\end{equation}
where $\epsilon_i=+1$ or $-1$ denotes the time-reversal parity of the corresponding variable.\footnote{For the usual densities, e.g., energy and particle number, $\epsilon_i=+1$ ~\cite{Callen1985, groot2013non}.} 
These reciprocity relations describe how microscopic reversibility constrains macroscopic processes. 

\subsection{Fluctuating work in isolated systems}

Stochastic thermodynamics extends classical phenomenological thermodynamic concepts to small, fluctuating systems far from equilibrium. 
It defines core concepts like heat, work, and entropy production for single fluctuating trajectories, so it is instructive to build a description of individual trajectories and of their ensemble. 
However, in the following, we restrict ourselves to externally driven closed systems, and postpone a more complete treatment of stochastic thermodynamics to the Appendix~\ref{sec:app-stochastic-thermodynamics}. 
For a comprehensive review of classical stochastic thermodynamics, including open systems, see Refs.~\cite{Jarzynski2011,Seifert2012,Falasco2025RMP}.

\paragraph{Hamiltonian dynamics.}
In classical mechanics, a closed, thermally isolated system that is potentially subject to external driving, with $N$ degrees of freedom, may be described by the canonical coordinates $\mathbf{q} = (q_1,\dots,q_N)$ and momenta $\mathbf{p} = (p_1,\dots,p_N)$. 
A microstate is specified by $\Gamma(\mathbf{p},\mathbf{q})$, a point in the $2N$-dimensional phase space, whose dynamics are generated by the Hamiltonian $H(\Gamma,\lambda)=H(\mathbf{q},\mathbf{p},\lambda)$, which may depend on time through an external control parameter $\lambda=\lambda(t)$ that is varied according to some prescribed protocol. 
Specifically, the system evolves according to Hamilton's equations of motion, 
\begin{align}
    \dot{q}_i = \frac{\partial H(\Gamma,\lambda)}{\partial p_i}, \qquad \text{and} \qquad
    \dot{p}_i = -\frac{\partial H(\Gamma,\lambda)}{\partial q_i},
\end{align}
where $i=1,\dots, N$ and the dot denotes the derivative with respect to time. 
The Liouville equation describes the evolution of the phase-space density $\mathcal{W}(\Gamma, t)$,
\begin{align} \label{eq:Liouville}
    \frac{\partial}{\partial t} \mathcal{W}(\Gamma,t) &= \{ H(\Gamma,\lambda), \mathcal{W}(\Gamma,t) \} 
    =\sum_{i=1}^N \left( \frac{\partial H(\Gamma,\lambda)}{\partial q_i} \frac{\partial \mathcal{W}(\Gamma,t)}{\partial p_i} - \frac{\partial H(\Gamma,\lambda)}{\partial p_i} \frac{\partial \mathcal{W}(\Gamma,t)}{\partial q_i} \right), 
\end{align}
where $\{\cdot,\cdot\}$ denotes the Poisson bracket. 
For a single particle in one dimension with Hamiltonian $H(x,p,\lambda) = \frac{p^2}{2m}+V(x,\lambda)$, where $m$ is the mass, $p=mv$ is the momentum, and $V(x,\lambda)$ is the potential, the Liouville equation~\eqref{eq:Liouville} coincides with the Klein--Kramers equation in the absence of friction, (thermal) noise, and non-conservative forces. 

Thermodynamically speaking, the simplest setting is a closed system that is potentially subject to external driving but thermally isolated from its environment. 
In practice, there will typically be some ultraweak coupling between the system and the environment, which enables the system to equilibrate at the inverse temperature $\beta$ of the environment for a fixed work parameter $\lambda$. 
The dynamics can be approximated by Hamilton's equations of motion if we assume that the energy exchange with the environment is negligible during driving, i.e., if the timescale over which the work parameter varies is much shorter than the equilibration time.

\paragraph{Fluctuation theorems.}
Since thermal fluctuations govern the dynamics of small systems, thermodynamic quantities such as heat and work fluctuate between realizations of the same process. 
Fluctuation theorems quantify these fluctuations and relate probabilities associated with quantities in a thermodynamic process to their time-reversed complement. 
They hold when the initial states of the forward and reverse processes are in thermal equilibrium, but are particularly useful because they also apply to systems driven arbitrarily far from equilibrium.  
These relations extend linear response theory, which admits a time-dependent description of the response of a system to a small perturbation in terms of fluctuations of the unperturbed system~\cite{Andrieux2008PRL}, while recovering fluctuation-dissipation theorems close to equilibrium. 

Formally, a fluctuation theorem is a symmetry relation that constrains the probability distribution $\mathcal{P}(\Omega)$ of a dimensionless trajectory functional $\Omega$ (such as dissipated work or entropy production). 
The distribution $\mathcal{P}(\Omega)$ obeys an \textit{integral fluctuation theorem} if 
\begin{equation} \label{eq:integral-fluctuation-theorem}
    \la e^{-\Omega} \ra = \int d\Omega\, \mathcal{P}(\Omega) e^{-\Omega} = 1, 
\end{equation}
where convexity of the exponential (Jensen's inequality), $e^{-\langle x \rangle} \leq \langle e^{- x} \rangle$, implies 
\begin{equation}
    \langle \Omega \rangle \geq 0, 
\end{equation}
which can be read as a version of the Clausius inequality~\eqref{eq:second-law} if the functional $\Omega$ corresponds to entropy production $\Sigma$. 
The \textit{detailed fluctuation theorem} is the stronger statement 
\begin{equation}
    \frac{\mathcal{P}(-\Omega)}{\mathcal{P}(\Omega)} = e^{-\Omega}, 
\end{equation}
which implies the integral counterpart \eqref{eq:integral-fluctuation-theorem}. 
Setting $\Omega=\sigma$, where $\sigma$ is the local entropy production \eqref{eq:sigma}, we observe that while negative entropy-production fluctuations are possible, they are exponentially suppressed relative to positive fluctuations of the same magnitude.   

The first such relation was observed by Evans and collaborators in simulations of sheared fluids in a non-equilibrium steady state~\cite{Evans1993PRL}, and later for transient trajectories starting from equilibrium~\cite{EvansSearles1994}. 
Gallavotti and Cohen provided the first rigorous proof of the fluctuation theorem in the setting of time-reversible, deterministic chaotic dynamics~\cite{GallavottiCohen1995PRL}. 
These results have since been extended to stochastic dynamics, including Langevin dynamics~\cite{Kurchan1998} and general Markov processes~\cite{Lebowitz1999}.

\paragraph{Classical Jarzynski equality.}
A central fluctuation theorem for thermodynamic work is the Jarzynski equality~\cite{Jarzynski1997PRL, Jarzynski1997PRE}, which relates free energy differences to work performed during non-equilibrium processes. 
Consider a process in which the system is initially prepared in thermal equilibrium at inverse temperature $\beta$ and parameter $\lambda_0=\lambda(0)$, which is then driven by some prescribed protocol to a final state at $\lambda(\tau)$ over time $\tau$. 
The Jarzynski equality states that the free-energy difference between the initial and final parameter values, $\Delta F \equiv F(\lambda_\tau) - F(\lambda_0)$, is related to the work performed on the system via 
\begin{equation} \label{eq:Jarzynski-equality}
    \la \exp{(-\beta W)} \ra  = \exp{(-\beta\Delta F)},
\end{equation}
which holds even when the driving takes the system arbitrarily far from equilibrium. 

Deriving the Jarzynski equality~\eqref{eq:Jarzynski-equality} for Hamiltonian dynamics is a simple exercise~\cite{Jarzynski1997PRL}. 
Consider the dynamics of an initial microstate $\Gamma_0$ sampled from the equilibrium Boltzmann-Gibbs distribution,
\begin{equation} \label{eq:inegral-fluctuation-theorem}
\mathcal{W}_{\lambda_0}^\mathrm{eq}(\Gamma_{0}) = \frac{1}{Z_0}\exp{(-\beta H(\Gamma_0,\lambda_0))}, 
\end{equation}
where $Z_t$ is the partition function, 
\begin{equation}
    Z_{t} = \int d\Gamma_t \exp{(-\beta H(\Gamma_t,\lambda_t))}, 
\end{equation}
which is related to the Helmholtz free energy $F_t$ via $\beta F_t = -\ln{Z_t}$ at time $t$. 
During the thermodynamic process, the system is thermally isolated and externally driven, so that the work performed on the initial microstate $\Gamma_0$ during a single realization over time $\tau$ corresponds to a change in the Hamiltonian, 
\begin{equation}
    W = H(\Gamma_\tau(\Gamma_0), \lambda_\tau) - H(\Gamma_0, \lambda_0), 
\end{equation}
where $\Gamma_\tau(\Gamma_0)$ is the time-evolved microstate. 

For Hamiltonian dynamics, 
\begin{equation}
\begin{split}
    \la \exp{(-\beta W)} \ra &= \int d\Gamma_0 \mathcal{W}_{\lambda_0}^\mathrm{eq} \exp{(-\beta W(\Gamma_0))}  \\
    &= \frac{1}{Z_0} \int d\Gamma_0 \exp{(-\beta H(\Gamma_\tau(\Gamma_0),\lambda_\tau))}  \\ 
    &=\frac{1}{Z_0} \int d\Gamma_\tau \exp{(-\beta H(\Gamma_\tau,\lambda(\tau))} 
    = \frac{Z_{\tau}}{Z_0} = \exp{(-\beta \Delta F)}, 
\end{split}
\end{equation}
where we have changed variables and used Liouville's theorem $\left|\det{({\partial \Gamma_\tau}/{\partial \Gamma_0})}\right|=1$.  

The Jarzynski equality~\eqref{eq:Jarzynski-equality} constrains the full work distribution, and remains valid for systems far from equilibrium. 
Furthermore, it holds in a wide range of classical systems; see Ref.~\cite{Jarzynski2011} and references contained therein. 

\section{Thermodynamic work in quantum systems}

Equipped with a classical understanding of thermodynamic phenomenology and a stochastic notion of work that leads to the fluctuation theorems, we can now move on to define work for quantum systems.

\subsection{Phenomenological quantum thermodynamics \label{sec:pheno}}

We begin the discussion by considering quasistatic processes during which a quantum system, $\mc{S}$, is always in equilibrium with a thermal environment, the heat bath $\mc{B}$. In complete analogy to the classical scenario, we assume that the Hamiltonian of the system, $H(\lambda)$, is parameterized by a control parameter $\lambda$. The parameter can be, e.g., the volume of a piston, the angular frequency of an oscillator, the strength of a magnetic field, etc.

Generally, the dynamics of $\mc{S}$ is then described by the Liouville-type equation $\dot{\rho}=L_{\lambda}(\rho)$, where the superoperator $L_{\lambda}$ reflects both the unitary dynamics generated by $H$ and the non-unitary contribution induced by the interaction with the environment. We further have to assume that the equation for the steady state, $L_{\lambda}(\rho^\mrm{ss})=0$, has a unique solution~\cite{Spohn1978} to avoid any ambiguities. 
As before, we will now be interested in thermodynamic state transformations, during which $\mc{S}$ remains in the equilibrium state corresponding to the instantaneous value of $\lambda$. 

In the limit of ultraweak coupling, the equilibrium state is given by the Gibbs state,
\begin{equation}
\label{eq:Gibbs_state}
\rho^\mrm{eq}=\e{-\beta H}/Z, \quad\text{where}\quad Z=\tr{\e{-\beta H}}\,,
\end{equation}
where, as before, $\beta$ is the inverse temperature of the environment. In this case, the thermodynamic entropy is given by the Gibbs entropy~\cite{Callen1985}, $ S^{\rm eq}=-\tr{\rho^\mrm{eq}\lo{\rho^\mrm{eq}}} = \beta \left(E - F \right)$, where $E=\tr{\rho^\mrm{eq}\, H}$ is the internal energy of the system, and $F=-1/\beta\,\lo{Z}$ denotes the equilibrium Helmholtz free energy. 

For isothermal, quasistatic processes, the change of thermodynamic entropy $d S^{\rm eq}$ can be written as
\begin{equation}
d S^{\rm eq}  = \beta\, \left( \tr{d\rho^\mrm{eq}\, H} + ( \tr{\rho^\mrm{eq}\, d H } - d F ) \right) = \beta\, \tr{d\rho^\mrm{eq}\, H}\,,
\end{equation}
where the second equality follows from simply evaluating terms. Therefore, two forms of energy can be identified~\cite{Alicki1979JPA,Quan2007PRE,Gemmer2009a,Deffner2019book}: heat is the change of internal energy associated with a change of entropy; work is the change of internal energy due to the change of an extensive parameter, i.e., change of the Hamiltonian of the system. Altogether, we have 
\begin{equation}
\label{eq:quantum_energy}
d E = \dbar Q + \dbar W \equiv \tr{d\rho^\mrm{eq}\, H} + \tr{\rho^\mrm{eq}\, d H}\,. 
\end{equation} 
The identification of heat $\dbar Q$ and work $\dbar W$ \eqref{eq:quantum_energy} is consistent with the second law of thermodynamics for 
quasistatic processes \eqref{eq:second-law} \emph{if}, and as will shortly see, \emph{only if} $\rho^\mrm{eq}$ is a Gibbs state \eqref{eq:Gibbs_state}.

It is worth emphasizing that for isothermal, quasistatic processes we further have
\begin{equation}
d S^{\rm eq}=\beta\, \dbar Q \quad \text{and}\quad d F= \dbar W\,,
\end{equation}
for which the first law of thermodynamics takes the form
\begin{equation}
d E=T\,d S^{\rm eq} +d F\,.
\end{equation}
In this particular formulation it becomes apparent that changes of the internal energy $d E$ can be separated into ``useful'' work, $d F$, and an additional contribution, $T\,d S^{\rm eq}$, reflecting the \emph{entropic cost} of the process.

\paragraph{Thermodynamics of non-Gibbsian equilibrium states.}

However, quantum systems can only be described by Gibbs states \eqref{eq:Gibbs_state} if they are at most ultraweakly coupled to the environment. Typically, quantum systems are correlated with their surroundings and interaction energies are \emph{not} negligible~\cite{Hu1992,Hanggi2008,Gelin2009}. For instance, it has been seen explicitly in the context of quantum Brownian motion~\cite{Horhammer2008,Bonanca2026BJP} that the system and environment are generically entangled. 

In such situations, the identification of heat only with changes of the state of the system \eqref{eq:quantum_energy} is no longer correct~\cite{Hanggi2008}. Rather, to formulate thermodynamics consistently, the energetic back action due to the correlation of system and environment has to be taken into account~\cite{Hanggi2008,Horhammer2008}. This means that during quasistatic processes, parts of the energy exchanged with the environment are not related to a change of the thermodynamic entropy of the system, but rather constitute the energetic price to maintain the non-Gibbsian state, i.e., coherence and correlations between system and environment. 

Denoting the non-Gibbsian equilibrium state by $\rho^\mrm{ss}$, we can write its von Neumann entropy $S^{\mrm{ss}}\equiv \tr{\rho^\mrm{ss}\lo{\rho^\mrm{ss}}}$ as
\begin{equation}
\begin{split}
 S^{\mrm{ss}} & = -\tr{\rho^\mrm{ss}\lo{\rho^\mrm{ss}}} + \left( \tr{\rho^\mrm{ss}\lo{\rho^\mrm{eq}}} - \tr{\rho^\mrm{ss}\lo{\rho^\mrm{eq}}} \right) \\
			     & = \beta \left[E -(F + T\, S(\rho^\mrm{ss} || \rho^\mrm{eq} )) \right]  = \beta\, ( \mc{E} -\mc{F} ),
	\end{split}
\end{equation}
where as before $E=\tr{\rho^\mrm{ss} H}$ is the internal energy of the system, and $\mc{F} \equiv F + T\, S(\rho^\mrm{ss} || \rho^\mrm{eq} )$ is the so-called the information
free energy~\cite{Sagawa2015}. Further, $S(\rho^\mrm{ss}|| \rho^\mrm{eq} )\equiv \tr{\rho^\mrm{ss}\left(\lo{\rho^\mrm{ss}} - \lo{\rho^\mrm{eq}}\right)}$ is the quantum relative entropy~\cite{Vedral2002}.

In complete analogy to the standard Gibbsian case \eqref{eq:quantum_energy}, we now consider isothermal, quasistatic processes, for which the infinitesimal change of the entropy reads
\begin{equation}
\begin{split}
		d S^{\mrm{ss}} & = \beta\, \left[ \tr{d\rho^\mrm{ss} H} + ( \tr{\rho^\mrm{ss} d H } - d\mc{F} ) \right] \\
				  & \equiv \beta\, (\dbar Q_\mrm{tot} - \dbar Q_{\rm c} ) \,,
\end{split}
\end{equation}
where we have identified the total heat as $\dbar Q_\mrm{tot}\equiv \tr{d\rho^\mrm{ss} H} $ and energetic price to maintain coherence and quantum correlations as $\dbar Q_\mrm{c}\equiv d\mc{F}-\tr{\rho^\mrm{ss} d H}$. 

The \emph{excess heat} $\dbar Q_\mrm{ex}$ is the only contribution that is associated with the entropic cost,
\begin{equation}
	d S^{\mrm{ss}} = \beta\, \dbar Q_{\rm ex}, \quad\text{and}\quad \dbar Q_{\rm ex} = \dbar Q_\mrm{tot} - \dbar Q_{\rm c}\,.
\end{equation}
Accordingly, the first law of thermodynamics takes the form
\begin{equation}
d E=\dbar W_\mrm{ex}+\dbar Q_\mrm{ex}\, ,
\end{equation}
where $\dbar W_\mrm{ex}\equiv\dbar W+\dbar Q_\mrm{c}$ is the excess work. Finally, Eq.~\eqref{eq:quantum_energy} generalizes for isothermal, quasistatic processes in generic quantum systems to
\begin{equation}
\label{eq:quantum_energy_ness}
d E= T\,d S^{\mrm{ss}}+d\mc{F}\,.
\end{equation}
It is worth emphasizing at this point once again that thermodynamics is a phenomenological theory, and as one expects, the fundamental relations hold for any physical system. Equation~\eqref{eq:quantum_energy_ness} has exactly the same form as Eq.~\eqref{eq:quantum_energy}; however, the ``symbols'' have to be interpreted differently when translating between the thermodynamic relations and the underlying statistical mechanics.

\paragraph{Application to quantum heat engines.}

The phenomenological approach has been instrumental in the study and development of quantum heat engines~\cite{Alicki1979JPA,Niedenzu2019quantum}. However, due to the large volume of literature on engines and other thermal devices, we refer the reader to recent comprehensive review articles, such as Ref.~\cite{Myers2022AVSqs,Arrachea2023RPP,Chand2026entropy,Cangemi2024PR}.

\subsection{Quantum work in Hamiltonian dynamics}
\label{sec:unitary_dynamics}

In complete analogy to the classical case, quantum stochastic thermodynamics is built upon fluctuation theorems. Conceptually, the most involved problem is how to identify heat and work for single realizations -- and even what a ``single realization'' means for a quantum system. Thus, it is \emph{a priori} not obvious how to distinguish heat and work in quantum dynamics. 

A significant body of literature circumvented this issue by considering only an \emph{isolated} quantum system that evolves under the time-dependent Schr\"odinger equation
\begin{equation}
i\hbar\,\pd_t \ket{\psi_t}=H_t\,\ket{\psi_t}\,.
\label{eq:schrodinger}
\end{equation}
However, as before, we are interested in describing thermodynamic processes that are induced by varying an external control parameter $\lambda_t$ during time $\tau$, so that $H_t=H(\lambda_t)$. In this case, the average work has to be identical to the change in expected energy,
\begin{equation}
\la W\ra=\la H_\tau\ra-\la H_0\ra\,,
\end{equation}
which can be interpreted as the simplest formulation of the first law of thermodynamics.

\subsubsection{Two-time measurement approach}
\label{sec:TTMA}

The most successful approach has become known as two-time energy measurement (TTM) approach~\cite{Kurchan2000,Tasaki2000,Talkner2007PRE,Talkner2007JPA,Talkner2008PRE,Campisi2011RMP,Campisi2011PT,Campisi2013NJPhys,Jarzynski2015PRX,Blattmann2017PRA}. Within this paradigm, quantum work is determined by the following protocol: At initial time $t=0$ a projective energy measurement is performed on the system; then the system is let to evolve under the time-dependent Schr\"odinger equation \eqref{eq:schrodinger}, before a second projective energy measurement is performed at $t=\tau$. 

Therefore, work $W$ becomes a stochastic variable, and for a single realization of this protocol we have
\begin{equation}
\label{eq:ttm_work}
W\left[\ket{n}; \ket{m}\right]=E_n'(\lambda_{\tau}) - E_m(\lambda_{0})\,,
\end{equation}
where $\ket{m}$ is the initial eigenstate with eigenenergy $E_{m}(\lambda_0)$ and $\ket{n} $ with $E_{n}'(\lambda_{\tau}) $ denotes the final state. 

The distribution of work values is then given by averaging over an ensemble of realizations of the same process,
\begin{equation}
\mc{P}(W)=\la \de{W-W\left[\ket{n}; \ket{m}\right]} \ra \,,
\end{equation}
which can be rewritten as
\begin{equation}
\label{eq:P_quwork}
\mathcal{P}(W)=\sumint_{m,n} \de{W-W\left[\ket{n};\ket{m}\right]}\,p\left(\ket{m}\rightarrow\ket{n}\right).
\end{equation}
In the latter equation, the symbol $\sumint$ denotes that we have to sum over the discrete part of the eigenvalue spectrum and integrate over the continuous part. Therefore, for systems with spectra that have both contributions, the work distribution will have a continuous part and delta-peaks; see, for instance, the Morse oscillator in Ref.~\cite{Leonard2015CP}.

Also in Eq.~\eqref{eq:P_quwork}, $p\left(\ket{m}\rightarrow\ket{n}\right)$ denotes the probability to observe a specific transition $\ket{m}\rightarrow\ket{n} $, 
\begin{equation}
p\left(\ket{m}\rightarrow\ket{n}\right)=\tr{\Pi_{n}\, U_{\tau}\, \Pi_{m}\,\rho^\mrm{eq}\,\Pi_{m}\, U_{\tau}^\dagger}\,,
\end{equation}
where $\rho^\mrm{eq}$ is the initial Gibbsian density operator \eqref{eq:Gibbs_state} of the system, $U_{\tau}$ is the unitary time evolution operator, $U_{\tau}=\mc{T}_> \e{-i/\hbar\,\int_0^{\tau}d t\,H_t}$, 
and $\Pi_{\nu}$ denotes the projector onto the eigenspace of the $\nu$th eigenvalue. 
For Hamiltonians with non-degenerate spectra, we simply have $\Pi_\nu=\ket{\nu}\bra{\nu}$. With these definitions, the transition probability is explicitly written as 
\begin{align}
\label{eq:P_transition}
p\left(\ket{m}\rightarrow\ket{n}\right)=\sum_{m,n}\bramatket{m}{\rho^{\mrm{eq}}}{m}\abs{\bramatket{n}{U_{\tau}}{m}}^2\,.
\end{align}

\paragraph{The quantum Jarzynski equality.}

It is then a relatively simple exercise to show that such a definition of quantum work fulfills a quantum version of the Jarzynski equality. To this end, we compute the average of the exponentiated work
\begin{equation}
\la \e{-\beta W}\ra=\int dW\,\mc{P}(W)\, \e{-\beta W}= \sumint_{m,n}  \e{\beta\,W\left[\ket{n};\ket{m}\right]}\,p\left(\ket{m}\rightarrow\ket{n}\right)
\end{equation}
Using the explicit expression for the transition probabilities \eqref{eq:P_transition} and for the 
Gibbs state \eqref{eq:Gibbs_state}, we immediately have
\begin{equation}
\label{eq:qu_Jarzynski}
\la \e{-\beta W}\ra=\e{-\beta \Delta F}\,.
\end{equation}
The latter theorem looks analogous to the classical Jarzynski equality \eqref{eq:Jarzynski-equality}. However, quantum work is a markedly different quantity than work in classical mechanics. It has been pointed out that work as defined from the two-time measurement is not a quantum observable in the usual sense, namely that there is no Hermitian operator whose eigenvalues are the classical work values~\cite{Talkner2007PRE}. The simple reason is that the final Hamiltonian does not necessarily commute with the initial Hamiltonian, $\com{H_\tau}{H_0}\neq 0$. Rather, quantum work is given by a time-ordered correlation function, which reflects that work is a thermodynamically non-exact, i.e., path-dependent, quantity.

As already mentioned, the two-time energy measurement approach has become one of the most successful and commonly accepted notions of quantum work. In part, this is due to its conceptual simplicity~\cite{Campisi2011RMP}, which also allows for other initial distributions, such as the microcanonical ensemble~\cite{Talkner2013NJP}. In addition, the approach has also been very successful in studying the thermodynamic properties of simple~\cite{Deffner2008PRE,Talkner2008PRE_work,Deffner2010CP,Quan2012PRE,Leonard2015CP,Yi2011PRE,Zon2008PRE,Yadalam2019PRA,Monnai2010PRE,Ford2012EJP,Santos2016NJP,Yi2013PRE_2,Liu2014PRE} as well as more complex quantum systems~\cite{Sotiriadis2013PRE,Smacchia2013PRE,Gong2014PRE,Fusco2014PRX,Garcia2017PRE,Lobejko2017PRE,Arrais2018PRE,Dong2019PRB,Nigro2019JSM,Cavalcante2026PRB,Fei2021PRB,Gu2022PRR,Lena2016PRA,Wang2018PRE,Kheirandish2020PLA,Dora2024PRB,Fei2019PRR,Fei2020PRL,Zuo2024PRE,Liu2023FP,Santini2023PRB}.

Moreover, the two-time measurement scheme has also been applied to $\mathcal{PT}$-symmetric and non-Hermitian quantum mechanics~\cite{Deffner2015PRL,Gardas2016SR,Erdamar2025PRR,Wei2018PRE,Zeng2017JPCO,Malakar2026PRB,Zhou2021PRE,Bose2026JPA,Reboiro2022QR}, relativistic quantum systems (in first quantization)~\cite{Deffner2015PRE}, and, in a field-theoretic setting, to driven quantum field theories~\cite{Bartolotta2018PRX,Costa2026PRD,Takeda2026PRD} and black holes~\cite{Pourhassan2021JHEP}. It should be noted, however, that projective measurements are somewhat ill-defined for quantum field theories, and alternative notions for quantum work are therefore required for a more consistent thermodynamic description~\cite{Ortega2019PRL,Teixido2020PRA,Pei2025PRL}.

Interestingly, the approach has also been extended to arbitrary other observables, which leads to generalized versions of the Jarzynski equality~\cite{Kafri2012PRA,Rastegin2013JSM,Rastegin2014PRE,NYH2017PRA,Watanabe2014PREa,Watanabe2014PREb,Prasanna2014NJP,Yi2013PRE,Rastegin2018IJTP,Funo2013PRE,Kwon2019PRX,Sone2023PRR}. Such generalized notions can be employed, e.g., to assess the performance of noisy quantum computers~\cite{Gardas2018SR}.

\subsubsection{Two-time measurement approach with ancilla \label{sec:ramsey}}

While the two-time measurement approach appears as a conceptually clean and experimentally well-motivated notion of quantum work, the actual realization in physical systems poses significant challenges. Some of the more successful attempts will be discussed below in Sec.~\ref{sec:experiments}. The main issue originates in the fact that, in order to determine the work distribution $\mathcal{P}(W) $ \eqref{eq:P_quwork}, projective measurements of the energy on a potentially complex quantum system are required.

This 
motivated the development of alternative schemes to determine the quantum work statistics. In particular, Refs.~\cite{Mazzola2013PRL} and~\cite{Dorner2013PRL} proposed Ramsey interferometry to measure the characteristic function of the work, $\chi(u)$. Here,  $\chi(u)$ is defined as the Fourier transform of $\mathcal{P}(W)$, and we can write
\begin{equation}
\begin{split}
\chi(u)&\equiv \int dW\,e^{i u W}\,\mathcal{P}(W)\\
&= \tr{U^\dagger_\tau\, e^{i u H_\tau}\,U_\tau\,e^{-i u H_0}\,\rho^{\mrm{eq}}(0)}\,,
\end{split}
\end{equation}
where as before $U_\tau$ is the time-evolution operator corresponding to Eq.~\eqref{eq:schrodinger}. Employing an ancilla qubit, Refs.~\cite{Mazzola2013PRL} and~\cite{Dorner2013PRL} then showed that the characteristic function $\chi(u)$ can be measured directly. As always, we assume that there is an external parameter $\lambda_t$, which drives the system with $H_0=H(\lambda_i)$ and $H_\tau=H(\lambda_f)$ according to some protocol $\lambda_0=\lambda_i$ and  $\lambda_\tau=\lambda_f$.

To this end, the total Hamiltonian is written as
\begin{equation}
H_\mrm{tot}(t)=\frac{\Delta}{2}\,\sigma_z\otimes \id_\mc{S} + \id_A\otimes H_\mc{S}+ h_\mrm{int}(t)\,,
\end{equation}
where $\frac{\Delta}{2}\,\sigma_z$ is the ancilla qubit, $H_\mc{S}$ is the \emph{time-independent} Hamiltonian of the system of interest $\mc{S}$, and $h_\mrm{int}(t)$ is a coupling term. This term contains the 
entire time dependence and reads
\begin{equation}
h_\mrm{int}(t)=\left(g_{\downarrow}(t) \ket{\downarrow}\bra{\downarrow}+g_{\uparrow}(t) \ket{\uparrow}\bra{\uparrow}\right) \otimes V\,.
\end{equation}
Here, $g_{\downarrow}(t)$ and $g_{\uparrow}(t)$ are external control parameters, and $V$ is a perturbation acting directly on the system of interest.

To measure the characteristic function $\chi(u)$, one then needs to follow the distinct Ramsey sequence of duration $\tau_R\geq\tau$:
\begin{itemize}
    \item[(i)] For times $t\leq0$, the qubit is decoupled from the system by holding the spin-dependent couplings fixed at $g_{\downarrow}(0)=g_{\uparrow}(0)=\lambda_i$. Moreover, the system is prepared in its Gibbs state, $\rho_\mc{S}^\mrm{eq}=\e{-\beta H_\mc{S}}/Z_0$, and the joint state is $\rho=\ket{\downarrow}\bra{\downarrow}\otimes\rho_\mc{S}^\mrm{eq}$.
    \item[(ii)] At $t=0$, a Hadamard operation is performed on the ancilla qubit.
    \item[(iii)] Then $g_{\downarrow}(t)$ and $g_{\uparrow}(t)$ are varied independently according to $g_{\downarrow}(t)=\lambda(t)$ for $t< \tau$ and $g_{\downarrow}(\tau)=\lambda_f$ for $\tau\leq t\leq \tau_R$; and $g_{\uparrow}(t)=\lambda_i$ for $t< \tau_R-\tau$ and $g_{\downarrow}(\tau)=\lambda(t-\tau_R+\tau)$ for $\tau_R-\tau\leq t\leq \tau_R$.
    \item[(iv)] Finally, at $t=\tau_R$, we have $g_{\downarrow}(\tau_R)=g_{\uparrow}(\tau_R)=\lambda_f$, which decouples the qubit and the system. The protocol is concluded with a final Hadamard on the qubit.
\end{itemize}

It is then easy to see~\cite{Dorner2013PRL} that the final state of the qubit can be written as
\begin{equation}
\label{eq:qubit}
\rho_A= \frac{1}{2}\left[\left(1+\mathfrak{R}(L_{\tau_R})\right) \ket{\downarrow}\bra{\downarrow}+i\,\mathfrak{I}(L_{\tau_R}) \ket{\downarrow}\bra{\uparrow}-i \mathfrak{I}(L_{\tau_R}) \ket{\uparrow}\bra{\downarrow}+\left(1-\mathfrak{R}(L_{\tau_R})\right) \ket{\uparrow}\bra{\uparrow}\right]\,,
\end{equation}
where $L_{\tau_R}$ is the decoherence factor. It can be expressed as
\begin{equation}
L_{\tau_R}=\ptr{S}{T_\uparrow^\dagger(\tau_R)\,T_\downarrow(\tau_R)\,\rho_\mc{S}^\mrm{eq}}\,,
\end{equation}
where 
\begin{equation}
\begin{split}
   T_\uparrow(\tau_R) & = \e{-i\Delta \tau_R/2}\,\e{-i(\tau_R-\tau) H(\lambda_f)}\,U_\tau, \\
   T_\downarrow(\tau_R)&= \e{i\Delta \tau_R/2}\,U_\tau\,\e{-i(\tau_R-\tau) H(\lambda_i)}\,.
\end{split}
\end{equation}
Finally, by identifying $\tau_R-\tau\equiv u$, the real and imaginary parts of $\chi(u)$ can be determined by performing quantum state tomography on the ancilla qubit, cf. Eq.~\eqref{eq:qubit}.

We note that the present discussion is a summary of the arguments proposed in Ref.~\cite{Dorner2013PRL}. However, Ref.~\cite{Mazzola2013PRL} introduced essentially the same idea. Reference~\cite{Talarico2016PRA} proposed an experimental implementation, and a related idea was also suggested in Ref.~\cite{Beyer2020PRR}, where the ancilla qubit is ``upgraded'' to an arbitrary control system, {on which work can be defined as an observable}. Also see Ref.~\cite{Campisi2013NJPhys} for an extension and implementation using a circuit QED setup, and Ref.~\cite{Mazzola2014IJQI} for another worked-out example system. 

\subsubsection{Two-time measurement approach with work meter}

The idea of obtaining the quantum work statistics \eqref{eq:P_quwork} with the help of ancillary systems was further formalized 
by introducing \emph{work meters}. To this end, 
the corresponding work can be determined from a positive operator-valued measurement (POVM)~\cite{roncaglia2014work}. Moreover, the ancilla qubit from above is ``upgraded'' to an entangling subsystem that serves as a meter that records the work performed on the other subsystem~\cite{cerisola2017using}.

To this end, consider again a finite-dimensional system described by a bipartite Hilbert space $\mathcal{H}=\mathcal{H}_{\mathcal{S}}\otimes\mathcal{H}_{\mathcal{A}}$, where the subscripts $\mathcal{S}$ and $\mathcal{A}$ represent the system and work meter that records the work performed on the system $\mathcal{H}_{\mathcal{S}}$. The total dynamics is described by the Schr\"{o}dinger equation \eqref{eq:schrodinger}. 

Again, the protocol starts with a product state 
\begin{align}
\ket{\psi}=\sum_{m}\frac{e^{-\beta E_m/2}}{\sqrt{Z_{0}}}\ket{m}_{\mathcal{S}}\otimes\ket{0}_{\mathcal{A}}\,.
\end{align}
Then, we let the system and work meter interact with each other, described by a unitary $\Gamma_{1}$,
\begin{align}
\Gamma_{1} = \sum_{\ell}\ketbrad{\ell}\otimes e^{iE_{\ell} P}\,,
\end{align}
where $P$ is a Hermitian operator yielding the translation
\begin{align}
    e^{i E_{\ell} P}\ket{E_{m}}=\ket{E_{m}-E_{\ell}}\,.
\end{align}
Therefore, $\Gamma_1$ is a controlled translation generated by $P$. The interaction encodes the energy difference into the displacement of the work meter. 
Note that this action of labeling enables one to encode the work information into the state of the work-meter system $\mathcal{H}_{\mathcal{A}}$ in a mathematically well-defined manner. 

After this interaction, we obtain
\begin{align}
\ket{\psi_0} = \Gamma_{1}\ket{\psi_{<0}}= \sum_{m} \frac{e^{-\beta E_m/2}}{\sqrt{Z_{0}}} 
\ket{m}_{\mathcal{S}}   \otimes   \ket{-E_m}_{\mathcal{A}}\,.
\label{eq:S+B_entangle_<0}
\end{align}
As we can see the reduced state of $\mathcal{H}_{\mathcal{S}}$ is the Gibbs state 
\begin{align}
    \text{tr}_{\mathcal{A}}\{\ketbrad{\psi_0}\}=\rho^{\mrm{eq}}=\frac{e^{-\beta H_0}}{Z_0}=\sum_{m}\frac{e^{-\beta E_m}}{Z_0}\ketbrad{m}\,.
\label{eq:initial_state}
\end{align}

To do work, we now consider a process in which $\ket{\psi_0}$ is evolved from $t=0$ to $t=\tau$ under $U_{t} \otimes \id_{\mathcal{A}}$.
The final state becomes 
\begin{align}
\ket{\psi_{\tau}}=\sum_{m}\frac{e^{-\beta E_m/2}}{\sqrt{Z_0}} U_{\tau}\ket{m}_{\mathcal{S}} \otimes  \ket{-E_m}_{\mathcal{A}}\,,
\label{eq:state_at_tau}
\end{align}
and the final reduced state of $\mathcal{H}_1$ is
\begin{align}
    \rho_{\tau}\equiv \text{tr}_{\mathcal{A}}\{\ketbrad{\psi_{\tau}}\}=U_{\tau}\rho_0U_{\tau}^{\dagger}\,.
\end{align}
Next, we let the system and the work meter interact with each other by applying a unitary
\begin{align}
    \Gamma_2 = \sum_{n}\ketbrad{n'}\otimes e^{-i E_n' P}\,.
\end{align}
Thus, after 
discretizing the displacement of the meter and representing the different work values as the eigenstates of the meter~\cite{Sone2026quantum}, the final state $\ket{\Psi_{\tau}}=\Gamma_2\ket{\psi_{\tau}}$ becomes
\begin{align}
\begin{split} 
\ket{\Psi_{\tau}}
&= \sum_{m,n}\frac{e^{-\beta E_m/2}}{\sqrt{Z_0}}\bramatket{n'}{U_{\tau}}{m}\ket{n'}_{\mathcal{S}}\otimes\ket{E_n'-E_m}_{\mathcal{A}}\\
&=\sum_{m,n}\frac{e^{-\beta E_m/2}}{\sqrt{Z_0}}\bramatket{n'}{U_{\tau}}{m}\ket{n'}_{\mathcal{S}}\otimes\ket{W_{nm}}_{\mathcal{A}}\\
&\approx\sum_{W}\left(\sum_{m,n}\frac{e^{-\beta E_m/2}}{\sqrt{Z_0}}\bramatket{n'}{U_{\tau}}{m}\delta_{W,W_{nm}}\ket{n'}_{\mathcal{S}}\right)\otimes\ket{W}_{\mathcal{A}}\\
&=\sum_{W}c_{W}\ket{\phi_{W}}_{\mathcal{S}}\otimes\ket{W}_{\mathcal{A}}\,,
\end{split}
\label{eq:TTM_env}
\end{align}
where we write 
\begin{align}
    W_{nm}\equiv W\left[\ket{n}; \ket{m}\right]\,.
\end{align}
Here, $\ket{W}$ again denotes the orthonormal basis labeled by the work $W$ satisfying $\braket{W}{W'}=\delta_{W,W'}$. 
In particular, $\ket{\phi_{W}}$ is 
\begin{align}
\ket{\phi_{W}}\equiv 
\frac{1}{c_{W}} \sum_{m,n}\frac{e^{-\beta E_m/2}}{\sqrt{Z_0}}\,
\bramatket{n'}{U_{\tau}}{m}\,
\delta_{W,W_{nm}} \ket{n'}\,.
\label{eq:phi_wTTM}
\end{align}
Since $\ket{\phi_{W}}$ has to be a pure state, by using $\braket{k'}{\ell'}=\delta_{k,\ell}$ and 
$\delta_{W,W_{nm}}^2=\delta_{W,W_{nm}}$, we obtain 
\begin{align}
  \abs{c_{W}}^2
  = \sum_{m,n} \frac{e^{-\beta E_m}}{Z_0}
\abs{\bramatket{n}{U_{\tau}}{m}}^2
    \delta_{W,W_{nm}} \,,
\end{align}
which is exactly the work distribution \eqref{eq:P_quwork}.

\subsubsection{Work operators}

However, as we have already emphasized in Sec.~\ref{sec:TTMA}, work is not a quantum observable in the usual sense, as there is no \emph{Hermitian} operator whose eigenvalues are the classical values. Nevertheless, it seems reasonably innocuous to define a possibly \emph{non-Hermitian} two-time work observable or a work operator as
\begin{equation}
    \hat{W}_{\tau} = U^{\dagger}_{\tau} H_{\tau} U_{\tau} - H_{0}\,,
    \label{eq:workop}
\end{equation}
where $ U^{\dagger}_{\tau} H_{\tau} U_{\tau}$ clearly represents the final Hamiltonian in the Heisenberg picture, and takes its eigenvalues as the possible outcomes of stochastic work in the quantum regime. 

Although not aiming at detailed fluctuation theorems or work statistics, this work operator was introduced for the first time by Bochkov and Kuzovlev~\cite{Bochkov1977}, in what is now called the exclusive viewpoint. Later, Allahverdyan and Nieuwenhuizen~\cite{Allahverdyan2005PRE} employed it in the inclusive viewpoint, showing that the Jarzynski equality could not be obtained from the average of its exponential. 

This result might seem at first glance very shocking, given the similarity of $\hat{W}_{\tau}$ and the two-time measurement approach. A question of principle often raised against the work operator (\ref{eq:workop}) is that work should characterize a process, rather than a state of the system, and consequently cannot be represented by an observable since a single projective measurement of it would determine its value~\cite{Talkner2007PRE}. 

Nevertheless, it has recently been argued in Ref.~\cite{Rolandi2026} that there is no such contradiction between being an observable and, at the same time, characterizing a process since $ U^{\dagger}_{\tau} H_{\tau} U_{\tau}$ is parametrized by the control protocol or process being performed. 
However, the issue is somewhat more subtle: as discussed in Ref.~\cite{Silva2024PRR}, the statistics $P(w) = \tr{M(w)\rho_\tau}$ of $\hat{W}_\tau$, with $M(w)$ given by
\begin{equation}
    M(w) = \sum_{j} \delta(w-w_{j}(\tau))\, |w_{j}(\tau)\rangle \langle w_{j}(\tau)| \,,
\end{equation}
and $w_{j}(\tau)$ and $\{ |w_{j}(\tau)\rangle\}$ denoting, respectively, the eigenvalues and eigenstates of $\hat{W}(\tau)$, furnish an average work that equals the change of the average energy, that is, $\la W \ra = \sum_{w}w P(w) = \tr{U^{\dagger}_{\tau} H_{\tau} U_{\tau} \rho_{0}} - \tr{H_{0} \rho_{0}}$, even when the initial state has coherences.

In addition, it has been claimed in Ref.~\cite{Rolandi2026} that within an autonomous description of a quantum thermodynamic setup, namely one in which there is no such thing as an external agent or work reservoir but only interacting quantum systems, the work operator (\ref{eq:workop}) naturally emerges as the quantity that provides the work statistics of a subsystem of interest. However, the Jarzynski equality cannot be recovered from it as expected.  

\subsubsection{One-time measurement approach}
\label{sec:OTM_closed}

The preceding discussion on work operators has further highlighted the crucial role that quantum measurements play in defining work. If the system $\mc{S}$ is initially prepared in a Gibbs state, the initial projective measurement of the energy is harmless. However, measuring energy at the end of the process might have significant thermodynamic consequences.

\paragraph{Neglected informational cost.}

Generally, the final state $\rho_\tau$ is a complicated non-equilibrium state. This means, in particular, that $\rho_\tau$ does not commute with the final Hamiltonian $H_\tau$, and one has to consider the back action on the system due to the projective measurement of the energy~\cite{Nielsen2010}. For a single measurement, $\Pi_{n}$, the post-measurement state is given by $\Pi_{n} \rho_\tau\Pi_{n}/p_n$, where $p_n=\tr{\Pi_{n}\,\rho_\tau}$. Thus, the system can be found on average in
\begin{equation}
\rho^M_\tau=\sum_{n}\Pi_{n}\,\rho_\tau\,\Pi_{n}\,.
\end{equation}
Accordingly, the final measurement of the energy is accompanied by a change of information, i.e., by a change  of the von Neumann entropy of the system, 
\begin{equation}
\Delta \mc{H}^M=-\tr{\rho^M_\tau\lo{\rho^M_\tau}}+\tr{\rho_\tau\lo{\rho_\tau}}\geq 0\,.
\end{equation}
Information, however, is physical~\cite{Landauer1991}, and its acquisition ``costs'' work.  This additional work has to be paid by the external observer, i.e., the measurement device. In a fully consistent thermodynamic framework, this cost should be taken into consideration. In fact, it was noted that neglecting this information cost is also inconsistent with the conservation of energy~\cite{Hovhannisyan2024Quantum}. 

\paragraph{Quantum work without measurements.}

To remedy this conceptual inconsistency arising from neglecting the informational contribution of the projective measurements, an alternative paradigm has been proposed~\cite{Deffner2016PRE}. For isolated systems, quantum work is clearly given by the change of internal energy. As a statement of the first law of thermodynamics, this holds true no matter whether the system is measured or not.

In fact, for thermal Gibbs states \eqref{eq:Gibbs_state}, measuring the energy is superfluous, since the state commutes with the Hamiltonian. 
Hence, an alternative notion of quantum work can be formulated that is fully based on the time-evolution of energy eigenstates. Quantum work for a single realization is then determined by considering how much the expectation value for a single energy eigenstate changes under the unitary evolution. Hence, we define
\begin{equation}
\label{eq:work_one}
\widetilde{W}_{m}\equiv \bra{m} U^\dagger_\tau\,H_\tau\,U_\tau\ket{m}-E_m\,.
\end{equation}
We can easily verify that the so-defined quantum work \eqref{eq:work_one} indeed fulfills the first law. To this end, we compute the average work $\la W\ra_{\widetilde{\mc{P}}}$,
\begin{equation}
\begin{split}
\la W\ra_{\widetilde{\mc{P}}}&=\sumint_{m} \bra{m} U^\dagger_\tau\,H_\tau\,U_\tau\ket{m}\,p_m-\tr{\rho^\mrm{eq}_0\, H_0}\\
&=\tr{\rho^\mrm{ss}_\tau\,H_\tau}-\tr{\rho^\mrm{eq}_0\,H_0}=\la W\ra \,,
\end{split}
\end{equation}
where $p_m=\e{-\beta\, E_m}/Z_0$ is the probability to find the system in the $m$th eigenstate at time $t=0$. 
Note that the average quantum work determined from two-time energy measurements is identical to the (expected) value given only knowledge from a single measurement at $t=0$. Most importantly, however, in this paradigm the external observer does not have to pay a thermodynamic cost associated with the change of information due to measurements.

\paragraph{Modified quantum Jarzynski equality.}

We have now seen that the first law of thermodynamics is immune to whether the energy of the system is measured or not, since projective measurements of the energy do not affect the internal energy. However, the informational content of the system of interest, i.e., the entropy, crucially depends on whether the system is measured. Therefore, we expect that the statements of the second law have to be modified to reflect the informational contribution~\cite{Deffner2013PRX}. In this paradigm, the modified quantum work distribution becomes
\begin{equation}
\label{eq:OTM}
\widetilde{\mc{P}}(W)=\sumint_{m} \delta(W-\widetilde{W}_{m})\, p_m\,,
\end{equation}
where as before $p_m=\e{-\beta\, E_m}/Z_0$. Now, we can compute the average exponentiated work,
\begin{equation}
\label{eq:Jarzynski_modified}
\la \e{-\beta W}\ra_{\widetilde{\mc{P}}}=\frac{1}{Z_0}\,\sumint_{m}\,\e{-\beta \,\bra{m} U^\dagger_\tau\,H_\tau\,U_\tau\ket{m}}\,.
\end{equation}
The right-hand side of Eq.~\eqref{eq:Jarzynski_modified} can be interpreted as the ratio of two partition functions, where $Z_0$ describes the initial thermal state. 
The second partition function,
\begin{equation}
\widetilde{Z}_\tau\equiv \sumint_{m}\,\e{-\beta \,\bra{m} U^\dagger_\tau\,H_\tau\,
U_\tau\ket{m}}, 
\end{equation}
corresponds to the best possible guess for a thermal state of the final system given only the time-evolved energy eigenbasis. 
This state can be written as
\begin{equation}
\widetilde{\rho}_\tau\equiv \frac{1}{\widetilde{Z}_\tau}\,\sumint_{m}\e{-\beta \,\bra{m} U^\dagger_\tau\,H_\tau\,U_\tau\ket{m}}\,U_\tau\ket{m}\bra{m}U^\dagger_\tau,
\end{equation}
which differs from the true thermal state, $\rho_\tau^\mrm{eq}=\e{-\beta H_\tau}/Z_\tau$.

As noted above, the ``quality'' of such a best possible guess in information theory is quantified by the relative entropy~\cite{Vedral2002}, which measures the distinguishability of two (quantum) states. Hence, consider
\begin{equation}
S(\widetilde{\rho}_\tau||\rho_\tau^\mrm{eq})=\tr{\widetilde{\rho}_\tau\lo{\widetilde{\rho}_\tau}}-\tr{\widetilde{\rho}_\tau\lo{\rho_\tau^\mrm{eq}}}\,,
\end{equation}
for which we compute both terms separately. For the first term, the negentropy of $\widetilde{\rho}_\tau$, we obtain,
\begin{equation}
\begin{split}
\tr{\widetilde{\rho}_\tau\lo{\widetilde{\rho}_\tau}}&=-\ln{(\widetilde{Z}_{\tau})}-\beta\, \tr{\widetilde{\rho}_\tau \sumint_{m} \bra{m} U^\dagger_\tau\,H_\tau\,U_\tau\ket{m}\,U_\tau\ket{m}\bra{m}U^\dagger_\tau }\\
&=-\ln{(\widetilde{Z}_{\tau})}-\beta \widetilde{E}_{\tau}\,,
\end{split}
\end{equation}
where we introduced the expected value of the energy, $\widetilde{E}_{\tau}$, under the time-evolved eigenstates,
\begin{equation}
\widetilde{E}_{\tau}=\frac{1}{\widetilde{Z}_{\tau}} \sumint_{m}\e{-\beta \,\bra{m} U^\dagger_\tau\,H_\tau\,U_\tau\ket{m}}\bra{m} U^\dagger_\tau\,H_\tau\,U_\tau\ket{m}\,.
\end{equation}
The second term, the cross entropy of $\widetilde{\rho}_\tau $ and $\rho_\tau^\mrm{eq} $, simplifies to
\begin{equation}
\begin{split}
\tr{\widetilde{\rho}_\tau\lo{\rho_\tau^\mrm{eq}}}&=-\lo{Z_{\tau}}-\beta \,\tr{\sumint_{m}\frac{1}{\widetilde Z_{\tau}} \e{-\beta \,\bra{m} U^\dagger_\tau\,H_\tau\,U_\tau\ket{m}} U_\tau\ket{m}\bra{m}U^\dagger_\tau\,H_\tau}\\
&=-\lo{Z_\tau}-\beta \widetilde{E}_{\tau}\,.
\end{split}
\end{equation}
Hence, the modified quantum Jarzynski equality \eqref{eq:Jarzynski_modified} becomes
\begin{equation}
\la \e{-\beta W}\ra_{\widetilde{\mc{P}}}=\e{-\beta \Delta F}\,\e{-S(\widetilde{\rho}_\tau||\rho_\tau^\mrm{eq})}\,,
\end{equation}
where as before $\Delta F=-1/\beta\, \lo{Z_\tau/Z_0}$. Jensen's inequality further implies
\begin{equation}
\label{eq:max_work}
\beta\,\la W\ra \geq \beta\,\Delta F + S(\widetilde{\rho}_\tau||\rho_\tau^\mrm{eq}),
\end{equation}
where we used $\la W\ra_{\widetilde{\mc{P}}}=\la W\ra$. 

By defining quantum work as an average over time-evolved eigenstates, we obtain a modified quantum Jarzynski equality \eqref{eq:Jarzynski_modified} and a generalized maximum-work theorem \eqref{eq:max_work}, in which the thermodynamic cost of projective measurements becomes apparent. These results become even more transparent by noting that similar versions of the maximum-work theorem have been derived in the thermodynamics of information~\cite{Sagawa2015}. In fact, it has proven useful to introduce the notion of an information free energy, 
\begin{equation}
\label{eq:infoF_var}
\widetilde{F}_\tau=F_\tau+S(\widetilde{\rho}_\tau||\rho_\tau^\mrm{eq})/\beta\,.
\end{equation}
Here, $\widetilde{F}_\tau$ accounts for the additional capacity of a thermodynamic system to perform work due to information~\cite{Deffner2013PRX}. Note that in Eq.~\eqref{eq:infoF_var}, $\widetilde{F}_\tau$ is computed for the fictitious equilibrium state $\widetilde{\rho}_\tau$.
We can rewrite Eq.~\eqref{eq:max_work} as
\begin{equation}
\beta\,\la W\ra \geq \beta\,\Delta \widetilde{F}\,.
\end{equation}
The latter inequality constitutes a sharper bound than the usual maximum-work theorem, and accounts for the extra free energy available to the system. Free energy, however, describes the usable, extractable work.

As we will see below, the one-time measurement (OTM) approach is also particularly convenient for generalizations to open system dynamics~\cite{sone2020quantum} and heat pumps~\cite{sone2023exchange}. Moreover, it also allows for classical analogs that quantify the thermodynamic consequences of inhomogeneous energy densities in phase space~\cite{sone2021jarzynski}. Even more importantly, it was also argued very recently that this one-time measurement approach is the only notion of quantum work that is fully consistent with the entanglement structure of quantum information theory~\cite{Sone2026quantum}.

\subsubsection{Endpoint measurement approach}

In the preceding section, we (at least implicitly) assumed that the quantum system is initially prepared in its Gibbs state \eqref{eq:Gibbs_state}. While this is certainly well-justifiable from a thermodynamic perspective, this assumption also hides a potential subtlety in defining work through measurements of the energy. Namely, by construction the Gibbs state and the Hamiltonian of the system commute, $\com{\rho^\mrm{eq}_0}{H_0}=0$, and hence the quantum system does not suffer from measurement back action. 

Generally, however, the quantum system might not be found in a Gibbs state at any time. This might be due to the fact that the system is strongly coupled to an environment or due to the presence of quantum coherences, cf. our discussion in Sec.~\ref{sec:pheno}. In this case, measuring the energy of the system becomes \emph{invasive}, which has led some to conclude that work can only be defined unambiguously if no quantum correlations are present~\cite{Perarnau-Llobet2017PRL,Silva2024PRR,Solinas2017PRA,Francica2024PRE}.

An alternative and more constructive proposal was put forth by Allahverdyan in Ref.~\cite{Allahverdyan2014Nonequilibrium}. In this work, it was suggested that if the initial measurement is inconsistent with thermodynamic notions of work, this measurement should simply be forgone; and then the quantum work distribution be constructed from only the available information. 

As before, we write initial and final Hamiltonians as
\begin{equation}
H_0=\sum_m E_m(\lambda_i) \,\Pi_m^0\quad\text{and}\quad H_\tau=\sum_n E_n'(\lambda_f) \,\Pi_n^\tau\,.
\end{equation}
To be consistent with the first law of thermodynamics, the first moment of the work needs to be given by the difference in expected energies, and hence we can generally write
\begin{equation}
\mc{P}_\mathrm{ep}(W)=\sumint_{m,n} \de{W-W\left[\ket{n};\ket{m}\right]}\,\mathfrak{p}_{n,m}, 
\end{equation}
where $W\left[\ket{n};\ket{m}\right]=E_n'(\lambda_f)-E_m(\lambda_i)$, as before.

However, in contrast to the two-time measurement approach, the probabilities $\mathfrak{p}_{n,m}$ are not determined by the unitary transitions, but rather constructed from the known initial and final Hamiltonians. Ref.~\cite{Allahverdyan2014Nonequilibrium} posits that a reasonable choice is given by
\begin{equation}
\mathfrak{p}_{n,m}\equiv \mathfrak{R}\left[\tr{\rho\,\Pi_m^0\,\Pi_n^\tau}\right]\,,
\end{equation}
where $\rho$ is the (measured) quantum state of the system, which is not necessarily diagonal in energy. Interestingly, $\mathfrak{p}_{n,m}$ can assume negative values, which means that here the quantum work statistics is governed by a quasiprobability distribution. Moreover, Ref.~\cite{Allahverdyan2014Nonequilibrium} explains that the $\mathfrak{p}_{n,m}$ can be determined from measurements only at the end of a thermodynamic process, and that $\mathfrak{p}_{n,m}$ are an optimal choice for the generalized conditional probabilities. However, it should also be obvious that such a construction is \emph{independent} of the specific protocol $\lambda_t$ provided that $\rho$ is known, which might not be as insightful as one desires. Nevertheless, Ref.~\cite{Allahverdyan2014Nonequilibrium} initiated the development of a more comprehensive paradigm built upon quasiprobability distributions.

{The above-described endpoint measurement approach is distinct from the endpoint scheme that was later proposed involving a single projective measurement at the end of the evolution or drive. 
The latter preserves initial coherences in the energy eigenbasis in exchange for uncertainty in the initial energy~\cite{Gherardini2021PRA, Mondkar2026PRA}.}

\subsubsection{Quasiprobability approach}

While there are many different ways to construct quasiprobability distributions, we will focus the present discussion on the so-called $q$-class quasiprobability~\cite{francica2022class,francica2022most}, which generalizes the Kirkwood-Dirac (KD)~\cite{Kirkwood1933Quantum,lostaglio2023kirkwood,Gherardini2024Quasiprobabilities} and the Margenau-Hill (MH)~\cite{Margenau1961Correlation,Allahverdyan2014Nonequilibrium,GarciaDiaz2020Quantum,Lostaglio2018Quantum} quasiprobability distribution of quantum work. In the following, we follow Refs.~\cite{francica2022class,francica2022most}. 

The $q$-class quasiprobability distributions are defined as 
\begin{align}
\begin{split}
\mathcal{P}_q(W)\equiv \sum_{i,j,k}&\mathfrak{R}\left[\bramatket{E_i}{\rho}{E_j}\bramatket{E_j}{U_{\tau}^{\dagger}}{E_k'}\bramatket{E_k'}{U_{\tau}}{E_i}\right]\\
&\times\delta\left(W-\left(E_k'-q E_i-(1-q)E_j\right)\right)\,.
\end{split}
\end{align}
where $q=0,1$ and $q=1/2$ recover the MH and FC cases. 
For a general initial state $\rho$, which 
need not commute with $H_{0}$ or $H_{\tau}$, the $n$-th moment of $W$ is given by~\cite{francica2022class} 
\begin{align}
    \ave{W^n}\equiv\int \mathcal{P}_q(W)W^ndW=(-i)^n\lim_{u\to 0}\partial_u^n
\chi_q(u)\,,
\end{align}
where, as before, $\chi_q(u)$ is the characteristic function of $\mathcal{P}_q(W)$ defined as 
\begin{align}
\begin{split}
\chi_q(u)
\equiv& \int \mathcal{P}_q(W)e^{iuW}dW\\
=&\frac{1}{2}\tr{e^{-iuqH_{0}}\rho e^{-iu(1-q)H_{0}}e^{iuU_{\tau}^{\dagger}H_{\tau}U_{\tau}}}+\frac{1}{2}\tr{e^{-iu(1-q)H_{0}}\rho e^{-iuq H_{0}}e^{iu U_{\tau}^{\dagger}H_{\tau}U_{\tau}}}\,.
\end{split}
\end{align}
Consistent with the first law, we have
\begin{align}
\ave{W}_{q}=\tr{H_{\tau}U_{\tau}\rho U_{\tau}^{\dagger}}-\tr{H_{0}\rho}\equiv \ave{W}\,.
\end{align}
However, for the quantum Jarzynski equality, we now obtain
\begin{align}
\begin{split}
    \la e^{-\beta (W-\Delta F)}\ra_{q}
    &\equiv\int \mathcal{P}_q(W)e^{-\beta W}dW=\Re\left[e^{\beta q H_{0}}\rho e^{-\beta q H_{0}}\left(\rho_{0}^{\text{eq}}\right)^{-1}U_{\tau}^{\dagger}\rho_{\tau}^{\text{eq}}U_{\tau}\right]\,.
\end{split}
\end{align}

Let $\Delta(\rho)=\sum_i \bramatket{E_i}{\rho}{E_i} \ketbrad{E_i}$ be the dephased initial state in the 
eigenbasis of $H_{0}$. Then, the relative entropy of coherence is~\cite{Baumgratz14}
\begin{align}
\ave{C}
=
S\left(\Delta(\rho)\right)-S(\rho),
\end{align}
where $S(\rho)=-\tr{\rho\ln\rho}$ denotes the von Neumann entropy. Writing the spectral decomposition of the initial state as $\rho=\sum_{\ell} r_{\ell}\ketbrad{r_{\ell}}$, 
one can introduce a stochastic coherence variable 
\begin{align}
C_{i\ell}
\equiv
\ln(r_{\ell})
-
\ln\bramatket{E_i}{\rho}{E_i}\,,
\end{align}
with a distribution
\begin{align}
\mc{P}(C)
=
\sum_{i,\ell}
r_{\ell}
|\braket{E_i}{r_{\ell}}|^2
\delta\left(
C+\ln\bramatket{E_{\ell}}{\rho}{E_{\ell}}-\ln r_{\ell}\right)\,.
\end{align}
With this, its average reproduces the relative entropy of coherence as
\begin{align}
\ave{C}\equiv\int dC\, \mc{P}(C)C
=
S\left(\Delta(\rho)\right)-S(\rho).
\end{align}
Moreover, the coherence variable satisfies a Jarzynski-like identity,
\begin{align}
\la e^{-C}\ra \equiv\int dC\, \mc{P}(C)e^{-C}=1.
\end{align}

We then define a joint quasiprobability distribution, 
\begin{align}
\begin{split}
\mathcal{P}_{q,q'}(W,C)=&\sum_{i,j,k,\ell}r_{\ell}\,\Re\left[\braket{E_i}{r_{\ell}}\braket{r_{\ell}}{E_j}\bramatket{E_j}{U_{\tau}^{\dagger}}{E_k'}\bramatket{E_k'}{U_{\tau}}{E_i}\right]\\
&\times \delta\left(W-E_k'+qE_i+(1-q)E_j\right)\\
&\times
\delta\left(C+q'\ln\bramatket{E_i}{\rho}{E_i}+(1-q')\ln\bramatket{E_j}{\rho}{E_j}-\ln r_{\ell}\right)\,.
\end{split}
\end{align}
The conventional two-time-measurement work distribution can be recovered through an exponential weighting of the coherence variable, 
\begin{align}
\int dC\,e^{-C}\mathcal{P}_{q,q'}(W,C)
=
\mc{P}(W).
\end{align}
Thus, the effect of initial coherence is encoded in the additional stochastic variable $C$.
Then, the resulting quantum Jarzynski equality becomes 
\begin{align}
\la
e^{-\beta(W-\Delta F)-C}\ra_{q,q'}
=
\tr{
\Delta(\rho)
\left(\rho_{0}^{\text{eq}}\right)^{-1}U_{\tau}^{\dagger}\rho_{\tau}^{\text{eq}}U_{\tau}}\,.
\label{eq:quasi-Jarzynski-equality}
\end{align}
When the initial state has thermal populations in the energy basis, $\bramatket{E_i}{\rho}{E_i}=e^{-\beta E_i}/Z_{0}$, 
Eq.~\eqref{eq:quasi-Jarzynski-equality} simply reduces to
\begin{align}
\la e^{-\beta(W-\Delta F)-C}\ra_{q,q'}
=1\,.
\end{align}
Hence, initial quantum coherence contributes an additional stochastic term $C$ to the usual Jarzynski equality. The corresponding second law can be obtained from Jensen's inequality, and we have
\begin{align}
\beta\left(\ave{W}-\Delta F\right)
+
\ave{C} = S(\rho_{\tau}||\rho_{\tau}^{\text{eq}})
\geq 0\,,
\end{align}
where $\rho_{\tau}=U_{\tau}\rho U_{\tau}^{\dagger}$ is the exact final state. Thus, the initial coherence $\ave{C}$ can be understood as an additional resource in the quantum work fluctuation theorem.

\subsubsection{Quantum ergotropy}

We conclude this part of the discussion with an even more detailed look at quantum coherence. To this end, we consider the \emph{quantum ergotropy}. Generally, the adjective ergotropic refers to the physiological mechanisms of a nervous system to favor an organism's capacity 
to expend energy~\cite{Gellhorn1970}. Generalizing this notion to physical systems, quantum ergotropy was then coined to denote the maximal amount of work that can be extracted by isentropic transformations~\cite{allahverdyan2004maximal}.

Mathematically, it is defined as 
\begin{align}
    \mathcal{E}(\rho)\equiv \tr{\rho H}-\min_{U\in\mathcal{U}(\mathcal{H})}\left[\tr{U\rho U^{\dagger}H}\right]\,,
\end{align}
where $\mathcal{U}(\mathcal{H})$ denotes the set of all unitaries acting on $\mathcal{H}$. 
When both $H$ and $\rho$ are nondegenerate, namely 
\begin{align}
    H = \sum_{i}E_i\ketbrad{E_i}\quad\text{with}\quad E_i\leq E_{i+1}
\end{align}
and 
\begin{align}
    \rho = \sum_{i} p_i\ketbrad{p_i}\quad\text{with}\quad p_{i}\geq p_{i+1}\,,
\end{align}
the optimal unitary is given by 
\begin{align}
U_{\text{opt}} = \sum_{i}\ketbra{E_i}{p_i}\,.
\end{align}
Therefore, the quantum ergotropy can be explicitly expressed as
\begin{align}
    \mathcal{E}(\rho) = \sum_{i,j}p_i E_j\abs{\braket{E_j}{p_i}}^2-\sum_{i}p_i E_i\,. 
\end{align}

The quantum ergotropy can also be written using the relative entropies~\cite{Sone21a,Francica20,Lobejko21}. Writing a second density matrix in its diagonal form, 
\begin{align}
\sigma = \sum_{i}s_i\ketbrad{s_i}\quad\text{with}\quad s_i\geq s_{i+1}, 
\end{align}
we have 
\begin{align}
\min_{U\in\mathcal{U}(\mathcal{H})}\left[S\left(U\rho U^{\dagger}||\sigma\right)\right] = \sum_{i}p_i\ln\left(\frac{p_i}{s_i}\right)=D(\rho||\sigma)\,,
\end{align}
where $D(\rho||\sigma)$ denotes the Kullback-Leibler (KL) divergence (or classical relative entropy). Letting $\rho^{\text{eq}}$ be a Gibbs state, 
\begin{align}
\rho^{\text{eq}}=\frac{e^{-\beta H}}{Z}=\sum_{i}\frac{e^{-\beta E_i}}{Z}\quad\text{with}\quad E_i\leq E_{i+1}\,, 
\end{align}
we have 
\begin{align}
\mathcal{E}(\rho) = \beta^{-1}\left(S(\rho||\rho^{\text{eq}})-D(\rho||\rho^{\text{eq}})\right)\,,  
\label{eq:rel_c}
\end{align}
which is overall independent of $\beta$, and holds for any $\beta$. 

The quantum ergotropy can also be separated into two terms: the incoherent ergotropy $\mathcal{E}_i(\rho)$ and the coherent ergotropy $\mathcal{E}_c(\rho)$~\cite{Francica20,Cakmak2020,Touil2021} as  
\begin{align}
\mathcal{E}(\rho)=\mathcal{E}_i(\rho)+\mc{E}_c(\rho)\,.
\end{align}
The incoherent ergotropy 
$\mathcal{E}_i(\rho)$ quantifies the maximum work that can be extracted from $\rho$ without manipulating its coherence. This is defined as~\cite{Francica20} 
\begin{align}
\mathcal{E}_i(\rho)\equiv \tr{(\rho-\tau)H}\,,
\end{align}
where $\tau$ is the called the coherence-invariant state of $\rho$, defined as  
\begin{align}
\tr{\tau H}=\min_{U\in\mc{U}^{(i)}}\tr{U\rho U^{\dagger}H}\,.
\end{align}
Here, $\mathcal{U}^{(i)}$ denotes the set of unitaries that do not change the coherence of $\rho$. 

On the other hand, the coherent ergotropy $\mathcal{E}_c(\rho)$ is the maximum work that is stored in the coherence, which is quantified by using the relative entropy of coherence~\cite{Baumgratz14}  
\begin{align}
\mc{C}(\rho)=S\left(\mathcal{D}(\rho)\right)-S(\rho)\,,
\end{align}
where $S(\rho)=-\tr{\rho\ln\rho}$ denotes the von Neumann entropy of $\rho$ and $\mathcal{D}$ denotes the purely dephasing map. Altogether, the coherent ergotropy can be rewritten as~\cite{Francica20,Sone21a}
\begin{align}
\label{eq:coh}
\beta\mathcal{E}_c(\rho)=\mathcal{C}(\rho)+S(\mathcal{D}(\tau)||\rho^\text{eq})-D(\rho||\rho^\text{eq})\,.
\end{align}
Even from this superficial discussion, it should be obvious that quantum ergotropy is an instrumental concept in building a consistent quantum thermodynamic theory~\cite{Choquehuanca2025PRA}.

\paragraph{From quantum work to quantum batteries.}

A particularly important application of the notion of quantum ergotropy is in the development of quantum batteries~\cite{Cenedese2026PRL,Du2025PRR,Morrone2023PRApp}. For instance, it has been recognized that two states of the same battery can have the same mean energy and still yield different amounts of work under the same discharge controls~\cite{allahverdyan2004maximal}. The work required by the load therefore provides a starting point for designing the charging protocol. Once the available discharge operations are known, the theory of quantum work can help identify which state to prepare.

One way to investigate this question is to use the model proposed in Ref.~\cite{Cakmak2020}. When several qubits are coupled to the same thermal bath, coherences can develop in the steady state and contribute to the work that can be extracted. To see when this is beneficial, one could compare the mean work recovered after storage when the qubits share a bath and when each qubit has its own bath. The bath temperature, storage time, and discharge controls would be the same in both cases, and the charging resources would be comparable. The choice of charging protocol also depends on how much the work delivered varies from one cycle to the next.

For a load that needs a minimum amount of work by a deadline, the fraction of cycles that meet this requirement gives a measure of reliability. For instance, Ref.~\cite{Mohan2026QuantumBattery} 
finds a trade-off between relative fluctuations in work and power during unitary charging. Extending their analysis to storage and discharge could show when charging faster makes the battery more likely to deliver the required work on time.

For a more comprehensive discussion of quantum batteries, we refer the reader to recent review articles on the topic~\cite{Campaioli2024RMP,Ferraro2026NRP}.

\subsection{Open system dynamics}

However, quantum systems are also (almost always) in interaction with some environment, with which the system exchanges energy in the form of heat. 
We now continue the discussion with a review of some of the more successful ways to extend notions of quantum work to open quantum dynamics. In the following, we thus consider a composite system $\mathcal{H}_{\mathcal{S}}\otimes\mathcal{H}_{\mathcal{B}}$. Here, $H_{\mathcal{S},t}$ is the system Hamiltonian, and $H_{\mathcal{B}}$ is the time-independent Hamiltonian of the bath. The total Hamiltonian becomes
\begin{align}
    H_{t}=H_{\mathcal{S},t}\otimes\id_{\mathcal{B}}+\id_\mathcal{S}\otimes H_{\mathcal{B}}+V_{t}\,,
\end{align}
where $\id$ denotes the identity operator and $V_{t}$ denotes the interaction Hamiltonian between the system and the bath. Here, we assume $V_0=0$.

As we outlined above in Sec.~\ref{sec:pheno}, in such open system dynamics any change of internal energy is comprised of heat and work. For systems initially prepared in Gibbs equilibrium states, we can then write
\begin{equation}
\label{eq:qu_first}
    d E = \dbar Q + \dbar W \equiv \tr{d\rho^\mrm{eq}\, H_{\mc{S}}} + \tr{\rho^\mrm{eq}\, d H_{\mc{S}}}\,.
\end{equation}
This poses the conceptual difficulty of having to experimentally distinguish heat and work.

The arguably simplest solution then is to measure the energy of the system, $\mc{S}$, and the bath, $\mc{B}$, individually. If $\mc{B}$ is a proper heat reservoir and in the ultra-weak coupling regime, its change of internal energy can be equated with the heat $E_\mc{B}=-\la Q\ra$. 
In such a paradigm, the two-time measurement approach to quantum work can be easily extended~\cite{Campisi2009PRL,Talkner2009JSM,Deffner2011PRL}, since work is then simply determined by subtracting the heat from the total change of the energy of $\mc{S}$. However, such a two-time measurement approach of the ``universe'' cannot be considered experimentally relevant, since it assumes that projective measurements on the entire heat bath are available.

\subsubsection{Quantum trajectories}

Generally, 
the heat bath is not accessible. A possible way to still distinguish heat and work was then proposed in Ref.~\cite{Alonso2016PRL}, where the authors assume that the system $\mc{S}$ can be continuously monitored by weak measurements~\cite{Nielsen2010}. For each run of a process, one can then assign a conditional quantum state $\tilde{\rho}_t$, which, when averaged over many realizations, reconstructs the actual quantum state, $\rho_t=\la\la \tilde{\rho}_t\ra\ra$. The evolution of $\tilde{\rho}_t$ is usually described by a stochastic master equation that contains a random parameter 
accounting for the detector shot noise.

Interestingly, any such stochastic master equation has a unitary part, which corresponds to the dynamics generated by $H_{\mathcal{S},t}$, and a non-unitary contribution that originates in the continuous interaction with the detector. Reference~\cite{Alonso2016PRL} then equates these two contributions to the heat and work of the quantum system, which closely mirrors Eq.~\eqref{eq:qu_first}. 

Similar constructions were also proposed for the entropy production along quantum jump trajectories~\cite{Horowitz2012PRE,Horowitz2013NJP,Liu2016PRE} and for stochastic wave functions~\cite{Leggio2013PRE,Leggio2013PRA}. Furthermore, Ref.~\cite{Suomela2015PRE} had a closer look at the fluctuations of work along quantum jump trajectories, and Ref.~\cite{Naghiloo2020PRL} measured heat and work. A related proposal defined work along trajectories in Bohmian mechanics~\cite{Sampaio2018PRA}, and finally Refs.~\cite{Micadei2020PRL,Micadei2021PRL,Micadei2023EPL} extended the idea of trajectory-based notions to determining quantum heat from Bayesian interference.

\subsubsection{Path integrals}

In classical stochastic thermodynamics, once our system of interest is found in permanent thermal contact with a heat bath, it has been shown that the phenomenological description of its effective dynamics is enough to derive its thermodynamics~\cite{PelitiBook,StrasbergPRE2017}. In other words, the microscopic details of the heat bath are irrelevant for this purpose, and rate equations, Fokker--Planck equations, or Langevin equations are all one needs to know. For quantum systems, this remains true only in the limit of strong damping~\cite{Deffner2011EPL}.

For general open quantum systems, the situation is more involved and constructing a consistent quantum stochastic thermodynamics requires special care. On the one hand, it is indeed correct that the master or Lindblad equations can be used to capture part of the thermodynamics of the system~\cite{Deffner2019book}. 
On the other hand, there is still an ongoing debate about how to obtain work and heat statistics for open quantum systems. 

An interesting way to approach this problem is by means of path integrals, as shown in Refs.~\cite{Quan2018PRL,Quan2018PRE}. The first idea is natural: assume that the system of interest plus its heat bath can be considered together as one Hamiltonian system evolving unitarily. Now, by means of an externally controlled parameter $\lambda_t$ that 
enters only 
the Hamiltonian of the system, one can perform work on this total thermally isolated system. The work statistics can then be formally written in terms of the characteristic function of the two-time measurement approach work distribution~\cite{Talkner2007PRE}, discussed in Sec.~\ref{sec:TTMA}, 
\begin{equation}
    \chi(u) = \int dW\,\mc{P}(W) e^{iu W} = \tr{U_\mrm{tot}(\tau)\,e^{-iu H_\mrm{tot}(\lambda_{0})}\rho(0)\, U^{\dagger}_{\mrm{tot}}(\tau)\,e^{iu H_\mrm{tot}(\lambda_{\tau})}}\,,
    \label{eq:chifunc}
\end{equation}
where $\mc{P}(W)$ is given by Eq.~(\ref{eq:P_quwork}), $H_\mrm{tot}(\lambda)$ denotes the Hamiltonian of the system plus the heat bath and their mutual interaction at a given value of $\lambda$, and $U_\mrm{tot}(\tau)$ is the time-evolution operator for the time-dependent global unitary dynamics under $\lambda_t$. 

As just mentioned, this is so far only a formal expression since $\mc{P}(W)$ requires measuring the eigenenergies of the \emph{whole} system. However, progress can be made by inserting the path-integral representation of the time-evolution operators and taking the partial trace over the heat bath degrees of freedom. This is feasible once a model for the heat bath is introduced. This model must be, at the same time, simple, to allow its exact partial trace, and complete, in the sense of describing all the relevant quantum non-equilibrium effects. In this respect, this minimal model plays the same role as the Langevin equation in classical stochastic thermodynamics. This is usually accomplished using the Caldeira-Leggett model. It describes the heat bath as a set of non-interacting harmonic oscillators, which allows its exact treatment due to the quadratic character of its respective action, and introduces a phenomenological quantity (the so-called spectral density of the harmonic modes) through which the dissipative regime can be modeled~\cite{Caldeira1983,Bonanca2026BJP}. 

Reference~\cite{Quan2018PRL} shows that the partial trace in Eq.~(\ref{eq:chifunc}) is feasible through the standard path-integral techniques used to treat quantum Brownian motion~\cite{Caldeira1983}. Therefore, $\chi(u)$ becomes dependent only on the effective time-dependent non-unitary dynamics of the system of interest, very much in the spirit of how classical stochastic thermodynamics makes use of Langevin equations. That is indeed remarkable since the work statistics of an open quantum system is obtained requiring, in a formal way, only the measurement of the total system energies. Nevertheless, there is a serious drawback: what measurements should we perform in the lab to compare with this predicted work statistics of the open system? Even if measurements of the system's energy were performed, we would not know how to split its value into work and heat contributions in order to compare with the theoretical work statistics just discussed. 
It then remains a central challenge in the field of quantum thermodynamics, both theoretically and experimentally, to split the change in the open system's energy into work and heat. 
A possible resolution was proposed in Ref.~\cite{Subasi2012PRE}, in which the path integral formalism is applied to decoherent histories; see in this context also Ref.~\cite{Miller2017NJP}.

\subsubsection{Conditional notions}

As noted above, the standard two-time measurement approach has several shortcomings for open systems: (i) it requires measurement of the inaccessible heat bath, and (ii) the second projective measurement destroys the correlations between the system and bath, which leads to the indistinguishability of heat and work from energy measurement on the system alone. To address these issues, Ref.~\cite{sone2020quantum} extends the one-time measurement approach to open quantum systems, and introduces the notion of conditional heat and work, which are distinguishable from the measurement on the system. 

We now assume that the initial state is of the product form $\rho_{\mathcal{S},0}^{\text{eq}}\otimes\rho_{\mathcal{B}}^{\text{eq}}$, where we define
\begin{align}
    \rho_{\mathcal{S},t}^{\text{eq}}=\frac{e^{-\beta H_{\mathcal{S},t}}}{Z_{\mathcal{S},t}}\quad\text{with}\quad Z_{\mathcal{S},t}=\tr{e^{-\beta H_{\mathcal{S},t}}}
\end{align}
as the Gibbs state defined by $H_{\mathcal{S},t}$ at time $t$, and 
\begin{align}
    \rho_{\mathcal{B}}^{\text{eq}}=\frac{e^{-\beta H_{\mathcal{B}}}}{Z_{\mathcal{B}}}\quad\text{with}\quad Z_{\mathcal{B}}=\tr{e^{-\beta H_{\mathcal{B}}}}
\end{align}
as the Gibbs state defined by $H_{\mathcal{B}}$. Here, $Z_{\mathcal{S}}(t)$ and $Z_{\mathcal{B}}$ are the partition functions. The total system evolves under the unitary as $U_{\mrm{tot}}(t)(\rho_{\mathcal{S},0}^{\text{eq}}\otimes\rho_{\mathcal{B}}^{\text{eq}})U_{\mrm{tot}}^{\dagger}(t)$.

In the one-time measurement approach, at time $t=0$, the energy of the system alone is measured, corresponding to the observable $H_{\mathcal{S},0}$. Suppose that the initial measurement outcome is $E_m$, which is the eigenvalue of $H_{\mathcal{S},0}$, with probability $e^{-\beta E_m}/Z_{\mathcal{S},0}$. 
After the measurement, the state is projected onto $\ket{E_m}$, i.e., the corresponding eigenstate of $H_{\mathcal{S},0}$. Then, we evolve the system until $t=\tau$. The reduced state of the system becomes
\begin{align}
 \Phi_{\tau}(\ketbrad{E_m}) = \text{tr}_{\mathcal{B}}\left\{U_{\mrm{tot}}(\tau)\left(\ketbrad{E_m}\otimes \rho_{\mathcal{B}}^{\text{eq}}\right)U^{\dagger}_{\mrm{tot}}(\tau)\right\}\,,   
\end{align}
where $\Phi_{\tau}$ denotes a completely positive trace-preserving (CPTP) map on $\mathcal{H}_{\mathcal{S}}$. 

Following the OTM scheme in Sec.~\ref{sec:OTM_closed}, we define the energy difference trajectory as 
\begin{align}
    \Delta \widetilde{E}_m = \tr{H_{\mathcal{S},\tau}\Phi_{\tau}(\ketbrad{E_m})}-E_m\,,
\end{align}
and the internal energy difference distribution as 
\begin{align}
    \widetilde{P}(\Delta E)=\sum_{m}\frac{e^{-\beta E_m}}{Z_{\mathcal{S},0}}\delta\left(\Delta E-\Delta \widetilde{E}_m\right)\,.
\end{align}
We 
find that the average of the internal energy difference with respect to the OTM distribution is exactly the average energy difference as
\begin{align}
\begin{split}
\langle \Delta E\rangle_{\widetilde{P}} 
&=\int d(\Delta E)\, \widetilde{P}(\Delta E)\Delta E=\tr{H_{\mathcal{S},\tau}\Phi_{\tau}\left(\rho_{\mathcal{S},0}^{\text{eq}}\right)}-\tr{H_{\mathcal{S},0}\rho_{\mathcal{S},0}^{\text{eq}}}=\langle \Delta E\rangle\,.
\end{split}
\end{align}
The corresponding Jarzynski equality becomes
\begin{align}
    \begin{split}
       \la e^{-\beta\Delta E}\ra_{\widetilde{P}} =\frac{1}{Z_{\mathcal{S},0}}\sum_{m}e^{-\beta\tr{H_{\mathcal{S},\tau}\Phi_{\tau}(\ketbrad{E_m})}}=\frac{\widetilde{Z}_{\mathcal{S},\tau}}{Z_{\mathcal{S},0}}\,,
    \end{split}
\end{align}
where we define
\begin{align}
\widetilde{Z}_{\mathcal{S},\tau}=\sum_{m}e^{-\beta\tr{H_{\mathcal{S},\tau}\Phi_{\tau}(\ketbrad{E_m})}}\,.
\end{align}
Accordingly, we can write 
\begin{align}
\langle e^{-\beta\Delta E}\rangle_{\widetilde{P}} = e^{-\beta\Delta F}\frac{\widetilde{Z}_{\mathcal{S},\tau}}{Z_{\mathcal{S},\tau}}\,.  
\end{align}

Now, let us define the following state, which is conditioned on the initial measurement outcomes, as 
\begin{align}
    \Theta_{\tau}=\sum_{m}\frac{e^{-\beta\tr{H_{\mathcal{S},\tau}\Phi_{\tau}(\ketbrad{E_m})}}}{\widetilde{Z}_{\mathcal{S},\tau}}U_{\mrm{tot}}(\tau)\left(\ketbrad{E_m}\otimes\rho_{\mathcal{B}}^{\text{eq}}\right)U^{\dagger}_{\mrm{tot}}(\tau)\,.
\end{align}
Then, considering the following quantum relative entropy 
\begin{align}
S\left(\Theta_{\tau}||\rho_{\mathcal{S},\tau}^{\text{eq}}\otimes\rho_{\mathcal{B}}^{\text{eq}}\right)
=\tr{\Theta_{\tau}\ln\Theta_{\tau}}-\tr{\Theta_{\tau}\ln\left(\rho_{\mathcal{S},\tau}^{\text{eq}}\otimes\rho_{\mathcal{B}}^{\text{eq}}\right)}\,,
\end{align}
we obtain 
\begin{align}    
S\left(\Theta_{\tau}||\rho_{\mathcal{S},\tau}^{\text{eq}}\otimes\rho_{\mathcal{B}}^{\text{eq}}\right)=-\ln\frac{\widetilde{Z}_{\mathcal{S},\tau}}{Z_{\mathcal{S},\tau}}-\beta\langle\widetilde{Q}\rangle_{\mathcal{B}}\,,
\end{align}
where 
\begin{align}
    \langle\widetilde{Q}\rangle_{\mathcal{B}} = \tr{H_{\mathcal{B}}\rho_{\mathcal{B}}^{\text{eq}}}-\tr{(\id_S\otimes H_{\mathcal{B}})\Theta_{\tau}}\,.
\end{align}
Here, $\langle\widetilde{Q}\rangle_{\mathcal{B}}$ represents the energy dissipation of the heat bath guessed from the initial measurement outcomes. Therefore, we can regard $\langle\widetilde{Q}\rangle_{\mathcal{B}}$ as a kind of heat, which we call conditional heat (or guessed heat).  

From the first law of thermodynamics, in a similar way, we can define the conditional work (or guessed work)~\cite{sone2021jarzynski},
\begin{align}
    \widetilde{W}=\Delta E-\langle \widetilde{Q}\rangle_{\mathcal{B}}\,.
\end{align}
Consequently, we obtain the following generalized Jarzynski equality
\begin{align}
    \la e^{-\beta\widetilde{W}}\ra_{\widetilde{P}}=e^{-\beta\Delta F}e^{-S\left(\Theta_{\tau}||\rho_{\mathcal{S},\tau}^{\text{eq}}\otimes\rho_{\mathcal{B}}^{\text{eq}}\right)}\,.
\end{align}

\section{Experimental implementations \label{sec:experiments}}

We conclude the discussion with a brief summary of experimental realizations of the most important notions of quantum work.

\subsection{Two-time measurement approach}

As we discussed above, the two-time measurement approach is particularly conducive for experimental implementations. Note, however, that for certain situations exponentially many experimental runs might be required~\cite{Jaramillo2017PRE}. Nevertheless, a first, practical proposal for ion traps was introduced in Ref.~\cite{Huber2008PRL}. Single ions in radio-frequency traps are quantum nanosystems with remarkable properties. They can be laser cooled to very low temperatures, reaching the motional ground state of the potential.  

To realize the two-time measurement approach, the following experimental set-up was considered~\cite{Huber2008PRL}: the ion trap consists of four blades, of which two are connected to a radio-frequency (rf) supply, and two are segmented with static (dc) voltages. The dc and rf blades are assembled perpendicular to each other. The complicated design of the trap is chosen to both define the trap potential with high accuracy and suppress precession of the trapped ion. 

\begin{figure}
\centering
\includegraphics[width=0.85\textwidth]{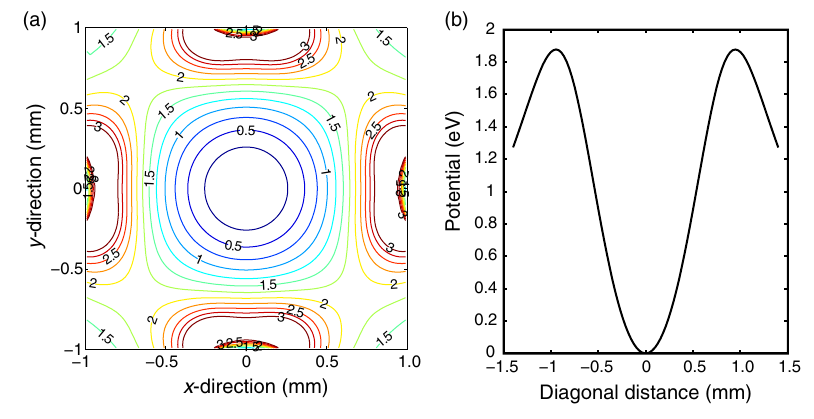}
\caption{(a) Contour plot of the potential in the ($x-y$)-plane in the experimental zone. (b) Cross-section trough potential along the $x=y$ direction. \textbf{[Credits: Figure from Ref.~\cite{Huber2008NJP}]} \label{fig:trap_potential}}
\end{figure}

Figure~\ref{fig:trap_potential} shows the results of numerical simulations for the confining potential. We observe that, to a very good approximation, the trap potential 
that of a harmonic oscillator. The time-dependence of the angular frequency is controlled externally by changing the (dc) voltage at the blades.  

In addition to the external motional degrees of freedom, the ion (e.g., $^{40}\mathrm{Ca}^+$) provides an internal, electronic level scheme. For the further exploration we choose a $\Lambda$ system comprising the ground state, $S_{1/2}$, and two exited states, $P_{1/2}$ and $D_{5/2}$.

The state $P_{1/2}$ has a short lifetime and decays rapidly into the $S_{1/2}$-state. This decay provides a high spontaneous photon scattering rate used for fluorescence detection. The $D_{5/2}$-state is populated with laser pulses if the spectral bandwidth of the $S_{1/2}-D_{5/2}$ exciting light field is small compared to the sideband structure. Thus, coherent laser pulses allow the exploitation and storage of the motional quantum state information in the internal quantum states.

The crucial point in experimental verifications of the quantum Jarzynski equality \eqref{eq:qu_Jarzynski} is the measurement of the instantaneous quantum state of the system. To this end, a measurement device filtering all possible states is necessary. 

\paragraph{Filtering scheme.}

The basic idea of the filtering is a null measurement~\cite{Huber2008PRL}. To this end, laser pulses are applied deterministically to depopulate a single state, which is supposed to be occupied. If a 
subsequent fluorescence measurement 
is dark, it can be concluded that the system was in the supposed state. 

The \textit{filtering scheme} starts with a sequence of laser pulses being applied to the ion on the narrow $S$ to $D$ transition, coherently processing its internal and external degrees of freedom. This pulse sequence is tailored such that the ion will end in the metastable $D_{5/2}$-state with certainty if the vibrational quantum state was $\ket{n_\mathrm{test}}$. Subsequently, the ion is illuminated with light resonant with the
$S_{1/2}$ to $P_{1/2}$ transition. If we observe no fluorescence, the ion is measured in the $D_{5/2}$-state. However, for vibrational states different from $\ket{n_\mathrm{test}}$, the laser pulse sequence leads to a superposition state, $\alpha \ket{S_{1/2}} + \beta \ket{D_{5/2}}$, such that there remains a non-vanishing probability $|\beta|^2$ of projecting the
superposition into $\ket{D_{5/2}}$ and thus observing no fluorescence. Therefore, the procedure is repeated a few times 
to ensure a sufficiently high quality of the filtering. Considering the evolution of the quantum state itself, the influence of the above sequence resembles the operating principle of a filter: its 
\textit{transmission} is unity for a certain input state $\ket{n_\mathrm{test}}$ and zero otherwise. This laser pulse sequence is adapted to reach all relevant eigenstates $\ket{m}$ and $\ket{n}$. Moreover, to estimate the time for one whole experiment cycle from preparation to detection, we can assume a few $10\upmu$s for sideband pulses and a few hundred $\upmu$s for fluorescence detection time. Thus, one cycle with multiple filtering iterations will take less than 10 msec, which is short compared to the lifetime of the $D_{5/2}$-state (1.2 sec for $^{40}\textrm{Ca}^+$)~\cite{Huber2008PRL}. Therefore, it is assured that the system is fairly isolated from the environment and the dynamics are, to very high accuracy, unitary. With 
this filtering scheme, the probability distribution of the work \eqref{eq:P_quwork} can be determined and, ultimately, the quantum Jarzynski equality \eqref{eq:qu_Jarzynski} can be verified.

\paragraph{Measurement protocol}

Now, we turn to the experimental {\it measurement protocol} of the work distribution. It can be summarized as consisting of four consecutive steps:
\begin{enumerate}
\item The trapped ion is prepared initially in a thermal state with mean phonon number $\la m\ra=(\exp(\beta\hbar \omega_0)-1)^{-1}$ in the electronic ground state 
by laser cooling and optical pumping. 
Specifically, the ion is laser cooled into the vibrational ground state, $\ket{m=0}$, and subsequently allowed to heat up for a certain time without laser cooling. As the heating rate of the ion within the trap can be precisely measured, this procedure is favorable for very low values of $\la m\ra$.
\item In the second step, the initial phonon number, $m$, is measured using the filtering scheme described in detail above. In this way, the initial energy eigenstate, $E_m$, is determined from spectroscopy measurements.
\item In the third step, the trap potential is varied from an initial value, $\omega_0$, to a final value, $\omega_\tau$. This changing potential will, in general, modify the motional state of the ion into a non-equilibrium state, while its internal electronic state remains unaffected.
\item In the last step, the new phonon number, $n$, is measured using the filtering scheme and the final energy eigenstate, $E_n'$, is determined. The distribution of the non-equilibrium work, $W= E_n'-E_m$ (cf. Eq.~\eqref{eq:ttm_work}), is then reconstructed by repeating the measurement sequence. By evaluating both sides of Eq.~\eqref{eq:qu_Jarzynski} for adiabatic and nonadiabatic processes, the Jarzynski equality can be verified.
\end{enumerate}

The above-explained \textit{filtering scheme} and \textit{measurement protocol} are described for the ideal case. However, in real experimental situations, one has to handle additional external perturbations. 
While the filtering and the measurement protocol remain unaffected, the outcome has to be analyzed carefully. 
In particular, the heating of the cold ion induced by the electronic surroundings might disturb the desired measurement.

Interestingly, Ref.~\cite{Huber2008PRL} reported only a theoretical proposal of the experimental set-up. Its implementation was published a few years later in Ref.~\cite{An2015NP} for ideal scenarios, and in Ref.~\cite{Smith2018NJP} for mildly open systems. In addition, Ref.~\cite{Batalhao2014PRL} also verified the quantum Jarzynski equality \eqref{eq:qu_Jarzynski}, but employing the Ramsey interferometry scheme outlined in Sec.~\ref{sec:ramsey}. Essentially, the same ion trap set-up was also employed to verify a quantum version of Landauer's principle~\cite{Yan2018PRL}, the corresponding fluctuation theorem~\cite{Yan2022Entropy}, and other information theoretic inequalities~\cite{Xiong2018PRL}. A related proposal was put forth in Ref.~\cite{Chiara2015NJP} to measure work \emph{and} heat in cold quantum gases, and Ref.~\cite{Kang2019SA} showed how to mitigate the measurement back action by performing collective measurements as proposed by Ref.~\cite{Perarnau-Llobet2017PRL}. 
 
\subsection{Work-meter model}
\label{sec:work-meter-experiment}

\begin{figure}
\centering
\includegraphics[width=1\columnwidth]{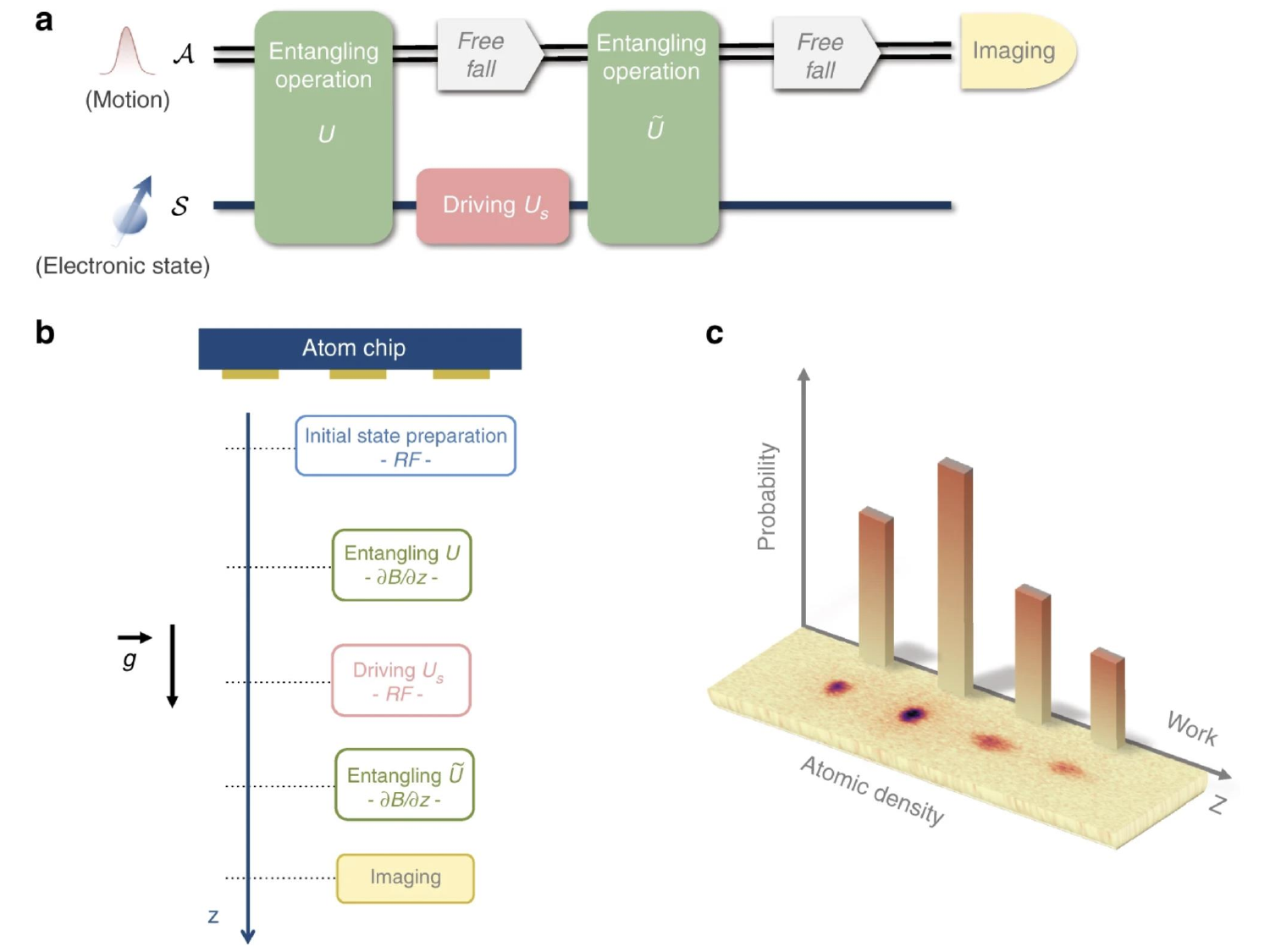}
\caption{The quantum work meter. Panel (a) gives the conceptual quantum work-meter circuit, while panel (b) shows its physical implementation on the atomic chip. Panel (c) is particularly intuitive. After the experimental sequence, four spatially separated atomic clouds appear, corresponding to the four possible values of work for the two-level system. The number of atoms in each cloud directly gives the probability of the corresponding work value. \textbf{[Credits: Figure from Ref.~\cite{cerisola2017using}]}}
\label{fig:atomic_chip}
\end{figure}

Reference~\cite{cerisola2017using} implemented the work-meter model on cold atomic ensembles to measure the work and samples the two-time measurement approach. The experiment used an ensemble of ultracold $^{87}\mathrm{Rb}$ atoms controlled by an atomic chip. The quantum system $\mathcal{S}$ is encoded in two Zeeman sublevels $\ket{1}=\ket{F=2, m_F=1}$ and $\ket{1}=\ket{F=2, m_F=2}$, which form an effective two-level system. The motional degree of freedom of the atoms, specifically their vertical motion, acts as the work meter $\mathcal{A}$, cf. Fig.~\ref{fig:atomic_chip}

The experiment begins with a Bose-Einstein condensate prepared in one Zeeman state. A radio-frequency 
(rf) pulse creates the desired populations of $\ket{1}$ and $\ket{2}$, which allows one to prepare different effective thermal states. The initial populations are chosen according to 
\begin{align}
    \beta E = \ln\frac{p_1}{p_2},
\end{align}
and the condensate is then released and allowed to fall freely. A magnetic-gradient pulse $\partial B/\partial z$ produces a state-dependent momentum kick. Since the force depends on the Zeeman state, this pulse entangles the internal state of the atom with its motional degree of freedom. Therefore, it realizes the first system-meter interaction $(U)$. An 
rf field then drives the internal two-level system. This implements the intermediate unitary $U_{\mathcal{S}}$, and its duration can be varied to generate different non-equilibrium processes. 

A second magnetic-gradient pulse $\partial B/\partial z$ with opposite direction and a different strength realizes another entangling gate $\widetilde{U}$. The two gradient pulses are calibrated in such a way that their relative momentum shifts reproduce the desired ratio of the initial and final energy splittings. The second pulse then completes the work-meter operation by encoding the energy difference in the atomic momentum. 

After a further time of flight, the different momentum components separate spatially. The experiment images the atomic cloud and determines both the position and atom number of each cloud. The position identifies the corresponding work value, while the relative number of atoms gives its probability. Thus, the complete work distribution is obtained directly from a single experimental realization.

The experimental system is especially simple because a two-level system has four possible combinations of initial and final energies. As a result, there are four possible work values: \begin{align}
    W_{nm}=E_n'-E_{m}\,.
\end{align}
These four values appear as four distinct atomic clouds. By changing the intermediate 
rf driving, one changes the transition probabilities 
$p\left(\ket{m}\rightarrow\ket{n}\right)\equiv\abs{\bramatket{n'}{U_{\mathcal{S}}}{m}}^2$. Therefore, one can change the relative populations of the four clouds, while the possible work values themselves remain fixed by the initial and final Hamiltonians.

This provides a particularly transparent realization of the quantum work distribution. The quantum work meter does not require reconstruction of $\mathcal{P}(W)$ from many measurements of a characteristic function. Instead, the work value is encoded directly into a physical degree of freedom, and the probability distribution is obtained by counting atoms. Reference~\cite{cerisola2017using} emphasizes that this makes the full statistical information about $\mathcal{P}(W)$ accessible in a single shot. 

The main application demonstrated in Ref.~\cite{cerisola2017using} is a test of the quantum Jarzynski equality~\eqref{eq:qu_Jarzynski}. 
The authors represent the allowed probability distributions as a ``Jarzynski manifold'' and show that the measurements performed with different driving protocols lie on the same manifold for a given temperature. They also extract the equilibrium free energy difference $\Delta F$ from the measured work distributions and find agreement with an independent calculation based on the partition functions.

\subsection{One-time measurement approach} 

\begin{figure}
\centering
\includegraphics[width=1\columnwidth]{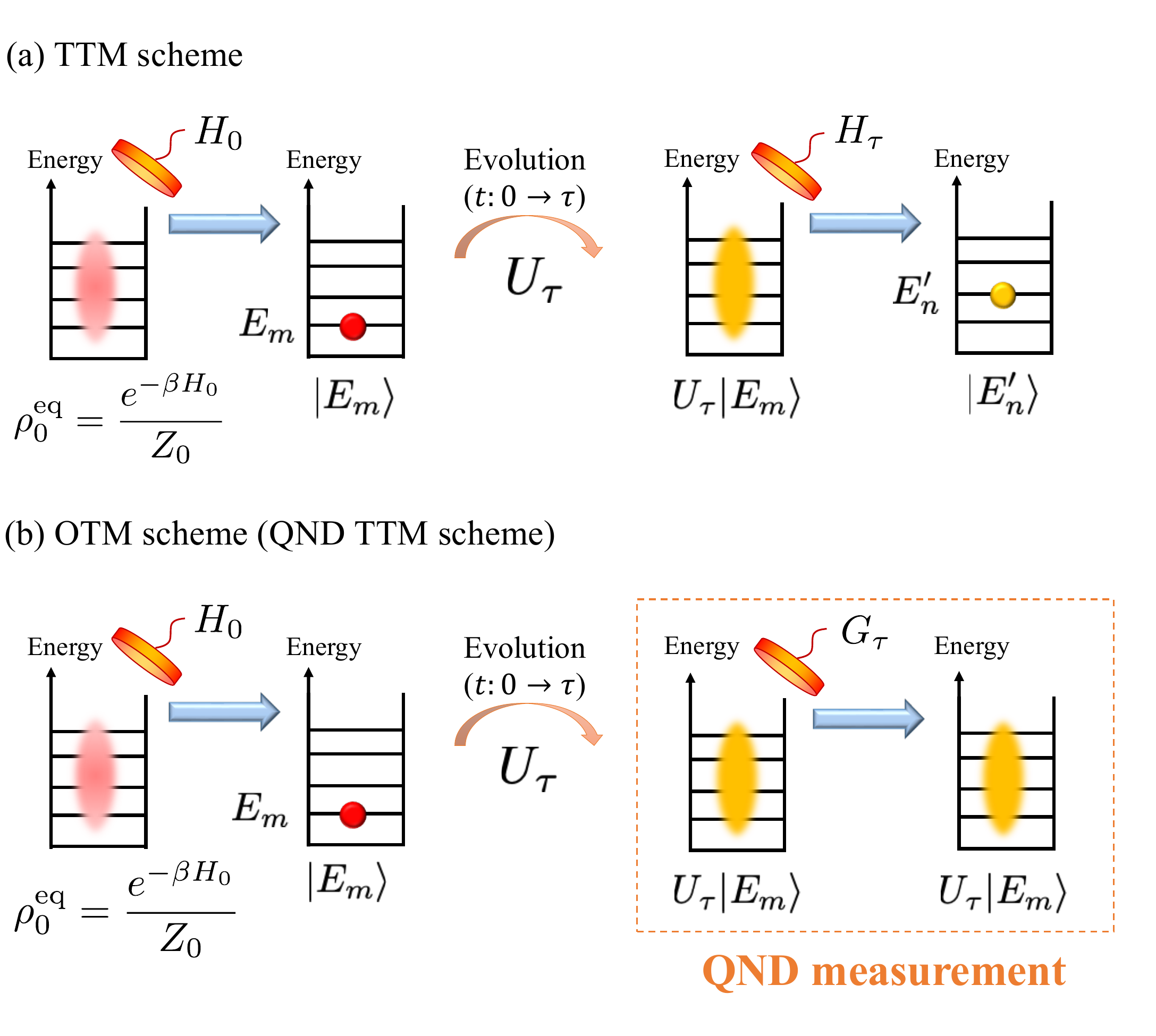}
\caption{Comparison between the TTM and OTM schemes. In (a) the standard TTM scheme, in which the second energy measurement is the final Hamiltonian, the projection measurement {projects} the evolved state $U_{\tau}|E_{m}\rangle$ onto $|E_n'\rangle$, the eigenstate of $H_{\tau}$. In (b), the OTM scheme, the final Hamiltonian $G_{\tau}\equiv \sum_{m=1}^{d} \bramatket{E_{m}}{U_{\tau}^{\dagger} H_{\tau}U_{\tau}}{E_{m}} U\ketbrad{E_{m}}U_{\tau}^{\dagger}$ is the conditional Hamiltonian with its pointer state $U|E_{m}\rangle$, which is equivalent to the evolved state conditioned on the initial measurement outcome. Therefore, this measurement {preserves} the state $U|E_{m}\rangle$. In this sense, this measurement is  {a QND measurement}. \textbf{[Credits: Figure 1 from Ref.~\cite{maeda2023detailed}]}} 
\label{fig:TTMvsOTM}
\end{figure}

The arguably second most important notion of quantum work, namely the one-time measurement approach, has also been realized experimentally. 
Concretely, consider an $n$-dimensional Hilbert space and again the work distribution in Eq.~\eqref{eq:OTM}, which we now rewrite as 
\begin{align}
\widetilde{\mc{P}}_f(W)=\sum_{m=1}^{d}\frac{e^{-\beta\bramatket{E_{m}}{H_0}{E_{m}}}}{Z_{0}}\abs{\bramatket{E_{m}}{U_{\tau}^{\dagger}U_{\tau}}{E_{m}}}^2\delta\left(W-\left(\bramatket{E_{m}}{U_{\tau}^{\dagger}H_{\tau}U_{\tau}}{E_{m}}-E_{m}\right)\right)\,,
\end{align}
where the index $f$ refers to a \emph{forward} realization of the process. Interestingly, this distribution can be understood as measuring the \emph{conditional} Hamiltonian
\begin{align}
    G_{\tau}=\sum_{m=1}^{d}\bramatket{E_{m}}{U_{\tau}^{\dagger}H_{\tau}U_{\tau}}{E_{m}}U_{\tau}\ketbrad{E_{m}}U_{\tau}^{\dagger}
\end{align}
at time $t=\tau$, which does not change the state $U_{\tau}\ket{E_{m}}$, cf. Fig.~\ref{fig:TTMvsOTM}).

Correspondingly, we can also construct a \emph{backward} work distribution
\begin{align}
    \widetilde{\mc{P}}_b(-W)=\sum_{m=1}^{d}\frac{e^{-\beta \bramatket{E_{m}}{U_{\tau}^{\dagger}H_{\tau}U_{\tau}}{E_{m}}}}{\widetilde{Z}_{\tau}}\delta(-W+\widetilde{W}_m)\,,
\end{align}
where the normalization factor is again $\widetilde{Z}_{\tau}=\tr{e^{-\beta G_{\tau}}}$. For the experimental implementation, we now consider the characteristic functions
\begin{align}
    \widetilde{\chi}_f(u)=\int dW\,\widetilde{\mc{P}}_f(W)e^{iuW}\quad\text{and}\quad
    \widetilde{\chi}_b(u)=\int dW\,\widetilde{\mc{P}}_b(-W)e^{-iuW}\,,
\end{align}
which again can be written using the Ramsey scheme~\cite{Dorner2013PRL,Mazzola2013PRL} as
\begin{align}
\widetilde{\chi}_f(u)=\tr{U_{\tau}^{\dagger}e^{iu G_{\tau}}U_{\tau}e^{-iu H_0}\rho_0^{\text{eq}}}\quad\text{and}\quad
\widetilde{\chi}_b(u)=\tr{U_{\tau}e^{iuH_0}U_{\tau}^{\dagger}e^{-iu G_{\tau}}\widetilde{\rho}_{\tau}}\,,
\end{align}
where here $\widetilde{\rho}_{\tau}=e^{-\beta G_{\tau}}/\widetilde{Z}_{\tau}$.
It is then easy to see that we have~\cite{maeda2023detailed}
\begin{align}
    \frac{\widetilde{\chi}_f(u)}{\widetilde{\chi}_{b}(-u+i\beta)}=\frac{\widetilde{Z}_{\tau}}{Z_{0}}=e^{-\beta\Delta F-S(\widetilde{\rho}_{\tau}||\rho_{\tau}^{\text{eq}})}\,.
\label{eq:symmetry_relation_OTM}
\end{align}

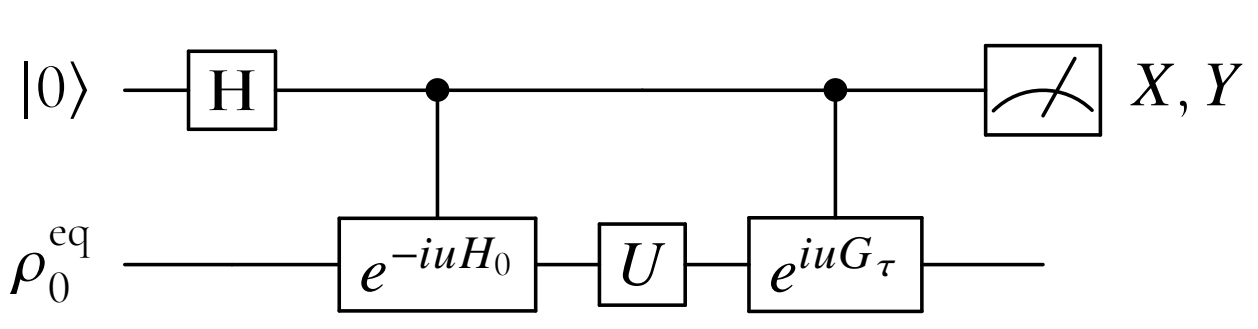
\begin{figure}[h]
\begin{center}
\mbox{
\Qcircuit @C=1em @R=1.3em {
\lstick{\ket{0}}& \gate{\mathbf{H}}& \ctrl{0} \qwx[1] &
\qw & \ctrl{0} \qwx[1] & \meter&~~~\text{$X$, $Y$} \\
\lstick{\rho_0^{\text{eq}}}&\qw & \gate{e^{-i u H_0}} & \gate{U}& \gate{e^{iu G_{\tau}}} &\qw \\
}
}
\end{center}
\caption{Quantum circuit for computing $\widetilde{\chi}_f(u)$. The expectation values of $X$ and $Y$ obtained by measuring the final state of the ancilla qubit are $\Re[\widetilde{\chi}_f(u)]$ and $\Im[\widetilde{\chi}_f(u)]$, respectively.  \textbf{[Credits: Figure 2 from Ref.~\cite{maeda2023detailed}]}}
\label{fig:ForwardCircuit}
\end{figure}

Both characteristic functions can be computed using quantum circuits. 
For the forward characteristic function, the quantum circuit is shown in Fig.~\ref{fig:ForwardCircuit}; the expectation values of $X$ and $Y$ are
\begin{align}
    \langle X\rangle = \Re\left[\widetilde{\chi}_f(u)\right]\, \quad\text{and}\quad
    \langle Y\rangle = \Im\left[\widetilde{\chi}_f(u)\right]\,.
\end{align}

\begin{figure}[h]
\begin{center}
\mbox{
\Qcircuit @C=1em @R=1.2em {
\lstick{\ket{0}}& \gate{\mathbf{H}}& \ctrl{0} \qwx[1] &
\ctrl{0} \qwx[1] & \qw & \ctrl{0} \qwx[1] & \ctrl{0} \qwx[1] &  \meter&~~~\text{$X$, $Y$} \\
\lstick{\widetilde{\rho}_{\tau}}&\qw & \gate{e^{iu G_{\tau}}}&\gate{\sigma_{\ell}} & \gate{U^{\dagger}}& \gate{e^{-iuH_{0}}}& \gate{\sigma_k} &\qw \\
}
}
\end{center}
\caption{{Quantum circuit for computing $F_{k\ell}$}: The expectation values of $X$ and $Y$ obtained by measuring the final state of the ancilla qubit are $\Re[F_{k\ell}]$ and $\Im[F_{k\ell}]$, respectively. \textbf{[Credits: Figure 3 from Ref.~\cite{maeda2023detailed}]}}
\label{fig:BackwardCircuit}
\end{figure}
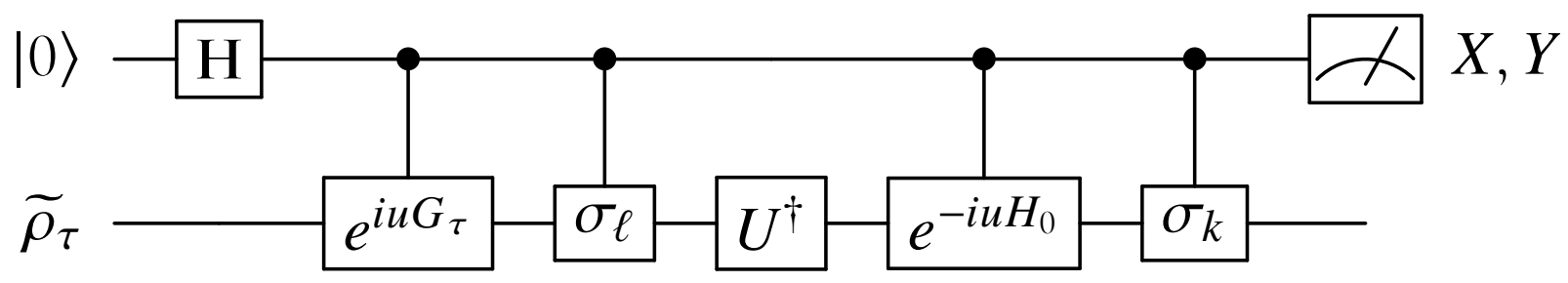

For the backward characteristic function $\chi_b(-u+i\beta)$, we need to be more careful. We have to first decompose $e^{\beta G_{\tau}}$ and $e^{-\beta H_0}$ 
in the basis of Pauli strings $\{\sigma_k\}_{k=1}^{d^2}$, where $\sigma_1=\id$ is the $d\times d$ identity matrix, and the Pauli strings satisfy
\begin{align}
    \frac{1}{d}\tr{\sigma_k\sigma_{\ell}}=\delta_{k\ell}=
    \begin{cases}
        0~~(k\neq \ell)\\
        1~~~(k=\ell)
    \end{cases}\,.
\end{align}
Then, we can write 
\begin{align}
    e^{-\beta H_0}=\sum_{k=1}^{d^2}\alpha_k^{(0)}\sigma_k\quad\text{and}\quad
    e^{\beta G_{\tau}}=\sum_{k=1}^{d^2}\alpha_k^{(\tau)}\sigma_k\,,
\end{align}
where
\begin{align}
    \alpha_k^{(0)}=\frac{1}{d}\tr{e^{-\beta H_0}\sigma_k}\quad\text{and}\quad
    \alpha_k^{(\tau)}=\frac{1}{d}\tr{e^{\beta G_{\tau}}\sigma_k}\,.
\end{align}
Here, the coefficients $\alpha_k^{(0)}$ and $\alpha_k^{(\tau)}$ are computable using a classical computer because we have full knowledge of $H_0$, $H_{\tau}$, and $U_{\tau}$. 
Altogether, we can write 
\begin{align}
    \widetilde{\chi}_b(-u+i\beta)=\sum_{k,\ell}\alpha_k^{(0)}\alpha_{\ell}^{(\tau)}F_{k\ell}\,,
\end{align}
where
\begin{align}
    F_{k\ell} = \tr{U_{\tau}\sigma_ke^{-iuH_0}U_{\tau}^{\dagger}\sigma_{\ell}e^{iu G_{\tau}}\widetilde{\rho}_{\tau}}\,,
\end{align}
which can be computed by using the 
quantum circuit illustrated in Fig.~\ref{fig:BackwardCircuit}. In this case, we have  
\begin{align}
\begin{split}
\langle X\rangle = \Re[F_{k\ell}]\quad\text{and}\quad
\langle Y\rangle = \Im[F_{k\ell}]\,.
\end{split}
\end{align}

\begin{figure}
\centering
\includegraphics[width=.6\columnwidth]{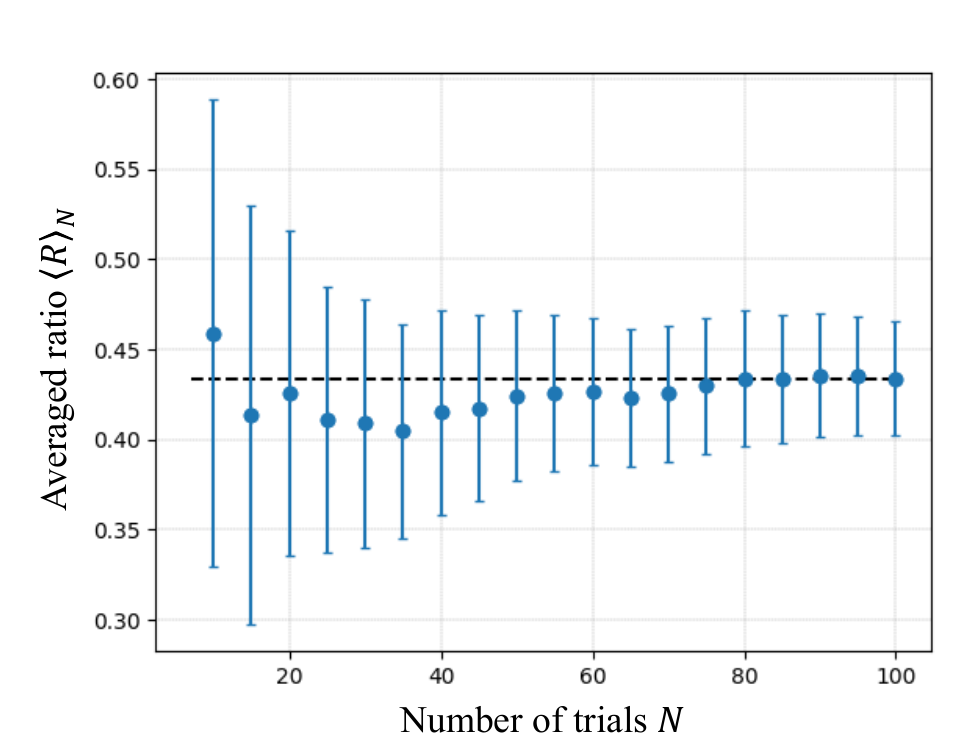}
\caption{Verification with IBM quantum computers. The dashed line is the exact value $R_{\text{true}}=0.433167$. The error bars represent $99\%$ confidence interval. As we increase the number of trials $N$, the averaged ratios $\ave{R}_N$ approach the exact value with a clear plateau starting from $N=75$. When $N=100$, we have $\ave{R}_{100}=0.433706$. \textbf{[Credits: Figure 4 from Ref.~\cite{maeda2023detailed}]}}
\label{fig:ExpResult}
\end{figure}

Reference~\cite{maeda2023detailed} employed the IBM cloud-based quantum computer for the verification of the symmetry relation in Eq.~\eqref{eq:symmetry_relation_OTM}. Their setup used $H_0=\omega_{z}(Z\otimes\id+\id\otimes Z)$ with the corresponding eigenbasis $\ket{E_1}=\begin{pmatrix}1&0&0&0\end{pmatrix}^{\top}$,  $\ket{E_2}=\begin{pmatrix}0&1&0&0\end{pmatrix}^{\top}$,  $\ket{E_3}=\begin{pmatrix}0&0&1&0\end{pmatrix}^{\top}$, and $\ket{E_4}=\begin{pmatrix}0&0&0&1\end{pmatrix}^{\top}$. The final Hamiltonian 
had the form $H_{\tau}=J(X\otimes X)$, and the unitary operator describing the evolution was set as $U_{\tau}=\exp\left(-i\frac{\omega_{y}}{2}\tau(Y\otimes\id+\id\otimes Y)\right)$. 
For the parameter values $\beta=0.5$, $\omega_{z}=2$, $\omega_{y}=2$, $J=1$, and $\tau=\pi/4$,  
the theoretical value follows as  
\begin{align}
R_{\text{true}}=\frac{\widetilde{\chi}_f(u)}{\widetilde{\chi}_b(-u+i\beta)}=0.433167\,,
\end{align}
for any $u$. 
Reference~\cite{maeda2023detailed} particularly focused on the $u=1$ case and determined $\langle X\rangle$ and $\langle Y\rangle$ for Figs.~\ref{fig:ForwardCircuit} and \ref{fig:BackwardCircuit} by performing 20,000 single-shot measurements, with median gate and qubit readout errors ranging between $10^{-4}$ and $10^{-2}$, and coherence times $T_1$ and $T_2$ ranging from around $17\mu\text{s}$ to $232\mu\text{s}$. By considering the absolute value of the ratio $ R\equiv \abs{\widetilde{\chi}_f(1)/\widetilde{\chi}_b(-1+0.5i)}$ and repeating the whole process $N$ times, one can compare $R_{\text{true}}$ and the average value $\langle R\rangle_N\equiv\frac{1}{N}\sum_{j=1}^{N}R_j$, where $R_j$ denotes the $R$ value at the $j$th trial. As $N$ increases, $\langle R\rangle_N$ approaches the true value as shown in Fig.~\ref{fig:ExpResult}, where the error bars represent the 99\% confidence interval. 

\section{Concluding remarks}

In this review, we have summarized and outlined the most important milestones in the development of a notion of \emph{quantum work}. Whereas for classical systems, work can be defined rather unambiguously, at large as well as small length scales, for deterministic as well as stochastic dynamics, as \emph{force times displacement}, finding a consistent and universally acceptable quantum notion has proven to be much harder. At its core, the issues originate in the rather invasive nature of quantum measurements~\cite{Beyer2020PRR,Chen2014PRE,Manzano2022AVSqs,Beyer2025PRL,Debarba2019NJP,Rubino2025npjQI,Touil2021entropy}. If and how to account for the thermodynamic cost of such measurements has given rise to a multitude of different proposals. 

The various notions covered in this review are summarized in Tab.~\ref{tab}.

\begin{table}[htbp]
    \centering
    \small
    \caption{Notions of and protocols for thermodynamic work in the quantum regime \label{tab}}
    \vspace{1ex}
    \begin{tabularx}{\textwidth}{@{} L{0.18\textwidth} Y L{0.3\textwidth} @{}}
        \toprule
        \textbf{Scheme} & \textbf{
        Operational idea} & \textbf{Key references} \\
        \midrule
        \midrule
        \multicolumn{3}{@{}l}{\textit{Two-time measurement (TTM) approach}} \\
        \midrule
        \midrule
        \addlinespace[2pt]
        Early formulations
        & Projective measurements of system energy at beginning and end of the evolution or drive of duration $\tau$.  
        First measurement destroys initial coherences between distinct energy eigenspaces. 
        & Proposed in~\cite{Kurchan2000,Tasaki2000,Talkner2007PRE}; trapped-ion proposal~\cite{Huber2008PRL}; using optical spectra~\cite{Heyl2012PRL}; trapped-ion experiment~\cite{An2015NP}; review~\cite{Talkner2016PRE}. \\ 
        \addlinespace 
        TTM with ancilla(s) 
        & Couple ancilla qubit to system and apply Ramsey sequence. 
        Perform quantum state tomography and Fourier transform to obtain work distribution. 
        & Interferometry~\cite{Dorner2013PRL, Mazzola2013PRL, Mazzola2014IJQI}; NMR experiment (verifies quantum Jarzynski equality)~\cite{Batalhao2014PRL}; open-system extension~\cite{Campisi2013NJPhys}; experimental proposal~\cite{Talarico2016PRA}; arbitrary control system~\cite{Beyer2020PRR}. \\ 
        \addlinespace
        TTM with work meter 
        & Single projective measurement of pointer after interaction with system before and after drive.
        Reproduces TTM statistics for ideal pointer. 
        & Proposal~\cite{roncaglia2014work}; cold-atom proposal~\cite{Chiara2015NJP}; cold-atom experiment (verifies quantum Jarzynski equality)~\cite{cerisola2017using}. \\ 
        \addlinespace
        \midrule
        \midrule
        \multicolumn{3}{@{}l}{\textit{Operator-based approaches}} \\
        \addlinespace[2pt]
        \midrule
        \midrule
        Work operator 
        & Work defined as Heisenberg-picture energy difference, whose expectation value gives the average work for any initial state. 
        In general, the initial and Heisenberg-evolved Hamiltonians do not commute, so eigenvalue statistics do not satisfy the Jarzynski equality. 
        & Exclusive work operator~\cite{Bochkov1977, Campisi2011PT}; inclusive work operator~\cite{Allahverdyan2005PRE, Engel2007EPL}; critique~\cite{Talkner2007PRE}; control-device observable~\cite{Beyer2020PRR}. \\
        \midrule
        \midrule
        \multicolumn{3}{@{}l}{\textit{One-time measurement (OTM) approach}} \\
        \midrule
        \midrule
        \addlinespace[2pt]
        Conditional-work scheme
        & Single projective measurement of system energy at beginning of drive of duration $\tau$.
        Work is expected final energy conditioned on outcome, minus initial energy.
        & Closed systems~\cite{Deffner2016PRE, sone2021jarzynski}; detailed fluctuation theorem and experiment~\cite{maeda2023detailed}; open-system extension~\cite{sone2020quantum}. \\ 
        \addlinespace
        Projective endpoint scheme
        & Single projective measurement of system energy at end of drive of duration $\tau$.
        Work inferred from energy statistics. 
        Preserves initial coherences in energy eigenbasis in exchange for uncertainty in initial energy.
        & Proposal and experiment~\cite{Gherardini2021PRA}; resource-resolved fluctuation theorems~\cite{Mondkar2026PRA}.\\ 
        \addlinespace
        Endpoint scheme
        & Work distribution constructed from available information, i.e., initial and final Hamiltonians, initial state, and evolution. 
        & Proposal~\cite{Allahverdyan2014Nonequilibrium}; 
        contextuality~\cite{Lostaglio2018Quantum}; comparison with TTM~\cite{GarciaDiaz2020Quantum}. \\
        \bottomrule 
    \end{tabularx}
\end{table}

\begin{table}[htbp]
    \ContinuedFloat
    \centering
    \small
    \caption[]{Notions of and protocols for thermodynamic work in the quantum regime (cont.)}
    \vspace{1ex}
    \begin{tabularx}{\textwidth}{@{} L{0.18\textwidth} Y L{0.3\textwidth} @{}}
        \toprule
        \textbf{Scheme} & \textbf{
        Operational idea} & \textbf{Key references} \\
        \midrule
        \midrule
        \multicolumn{3}{@{}l}{\textit{Quasiprobability schemes}} \\
        \addlinespace[2pt]
        \midrule
        \midrule
        Work quasiprobabilities
        & Replaces work distribution with (possibly negative or complex) quasiprobability over initial and final energies, whose first moment gives the average work for any state. 
        Retains initial coherence. 
        Reduces to TTM for states diagonal in initial energy eigenbasis. 
        Accessed via weak or sequential measurements, or interferometry. 
        & Kirkwood--Dirac (KD) quasiprobability~\cite{Kirkwood1933Quantum,lostaglio2023kirkwood,Gherardini2024Quasiprobabilities}; Margenau--Hill (MH) quasiprobability~\cite{Margenau1961Correlation,Allahverdyan2014Nonequilibrium,GarciaDiaz2020Quantum,Lostaglio2018Quantum}; general $q$-class of quasiprobabilities~\cite{francica2022class,francica2022most}.\\
        \addlinespace
        \midrule
        \midrule
        \multicolumn{3}{@{}l}{\textit{Ergotropy}} \\
        \addlinespace[2pt] 
        \midrule
        \midrule
        Ergotropy or extractable work
        & Defines the maximum work extractable by cyclic unitary driving. Protocol-independent. 
        Expressible as difference between quantum and classical relative entropies. 
        & Introduced in~\cite{allahverdyan2004maximal}; coherent and incoherent contributions~\cite{Francica20, Sone21a}; relative-entropy form and OTM scheme~\cite{Sone21a}. \\
        \addlinespace
        \midrule
        \midrule
        \multicolumn{3}{@{}l}{\textit{Open-system schemes}} \\
        \addlinespace[2pt]
        \midrule
        \midrule
        TTM                              
        & Work distribution of system plus bath, typically estimated via interferometry with ancillary system. 
        & Weak coupling~\cite{Campisi2009PRL}; arbitrary coupling~\cite{Talkner2009JSM}; interferometric proposals~\cite{Campisi2013NJPhys, Mazzola2014IJQI}; trapped-ion experiment with engineered decoherence~\cite{Smith2018NJP}; quantum-computer experiment~\cite{Oftelie2025QST}. 
        \\
        \addlinespace
        Work operator
        & Work measured on control device. Requires no knowledge of system Hamiltonian. 
        & Proposal and operational work fluctuation theorem~\cite{Beyer2025PRL}. 
        \\
        \addlinespace
        OTM                                
        & Guessed quantum work extracted from OTM scheme with initial measurement of system energy. 
        Modified Jarzynski equality and maximum-work principle. 
        & Introduced in~\cite{sone2020quantum}.\\
        \addlinespace
        Weak/continuous- measurement work 
        & Work inferred from 
        non-projective energy measurements, e.g., via continuous weak monitoring along individual quantum trajectories. 
        & Quantum jumps~\cite{Horowitz2012PRE, Hekking2013PRL, Suomela2015PRE}; individual trajectories~\cite{Alonso2016PRL};
        superconducting-qubit experiment~\cite{Naghiloo2020PRL}; modified fluctuation theorems~\cite{Watanabe2014PREa, Watanabe2014PREb, Talkner2016PRE}. \\
        \bottomrule
    \end{tabularx}
\end{table}

From the preceding discussion, it is apparent that properly defining quantum work has been a main pillar of research in \emph{quantum thermodynamics}. The present review is focused exclusively on the version of quantum work, but there are many other thermodynamic questions that are directly related. For instance, significant efforts have been made to formulate the second law of thermodynamics in the quantum regime~\cite{Deffner2010PRL,Deffner2013PRE,Brodier2020JPA}. Such formulations generally require a universal understanding of entropy production~\cite{Spohn1978JMP,Manzano2019PRL,Landi2012RMP,Moreira2025CQG,Strasberg2019PRE,Santiago2023npjQI,Aguilar2022PRA,Gherardini2018QST,Huang2023PRA} and its genuine contributions from quantum correlations~\cite{Santos2019NPJqi}. However, doing the developments of quantum entropy production justice is beyond the scope of the present review, since it would require a comprehensive discussion of, e.g., general open system dynamics~\cite{Manzano2018PRX,Chiara2022PRR,Li2024PRA} or quantum phase space~\cite{Deffner2013EPL,Santos2017PRL}.

\section{Acknowledgments}
A.S. acknowledges support from the U.S. NSF under Grant No.~OSI-2328774. A.S. also acknowledges support from PHY-2425180 and Cooperative Agreement PHY-2019786.
S.D. acknowledges support from the John Templeton Foundation under Grant No. 63626. This work was supported by the U.S. Department of Energy, Office of Basic Energy Sciences, Quantum Information Science program in Chemical Sciences, Geosciences, and Biosciences, under Award No. DE-SC0025997. The authors also acknowledge support from the Foundational Questions Institute (FQxI).

\section{Appendix: Stochastic thermodynamics for classical systems} 

\label{sec:app-stochastic-thermodynamics}
Here, we include a more detailed treatment of stochastic dynamics, energetics, and entropy for systems coupled to a thermal environment.  

\subsection{Stochastic dynamics}
The Langevin equation is a stochastic differential equation that describes how a system evolves under deterministic forces (such as friction and potential gradients) and fluctuating random forces (as in Brownian motion), whose solutions are single noisy trajectories. 
The same random process can be described by the Fokker--Planck equation, which is a partial differential equation describing the time evolution of the probability density function of the position (and/or velocity), and instead follows how the probability distribution of the state of the system evolves in time. 

\paragraph{General Langevin and Fokker--Planck equations.} 
In It\^{o} form, a Langevin equation has the general form 
\begin{equation} \label{eq:Langevin-general-form}
    dx=A(x,t)\,dt + B(x,t)\,d\mathbf{W}_t,
\end{equation}
where $x$ is the position, $A$ is the deterministic drift coefficient, $B$ is the noise amplitude, and $d\mathbf{W}_t$ is the increment of the Wiener process with $\langle d\mathbf{W}_t \rangle =0$ and $\langle (d\mathbf{W}_t)^2 \rangle =dt$. 
The corresponding probability density $\mathcal{P}=\mathcal{P}(x,t)$ obeys the Fokker--Planck equation 
\begin{equation} \label{eq:FP-general-form}
    \frac{\partial \mathcal{P}}{\partial t} = - \frac{\partial}{\partial x} \left[ A(x,t) \mathcal{P} \right] + \frac{1}{2} \frac{\partial^2}{\partial x^2} \left[ B(x,t)^2 \mathcal{P} \right], 
\end{equation}
where the drift term in the Langevin equation enters the first derivative in the Fokker--Planck equation, while half the noise amplitude squared is the diffusion coefficient.

Typically, physical systems are complex and multivariate. 
For several variables $\mathbf{z} = (z_1, \dots,z_n)$, the Langevin equation~\eqref{eq:Langevin-general-form} is modified to have the general form
\begin{equation} \label{eq:Langevin-general-form-multi}
    d\mathbf{z} = \mathbf{A}(\mathbf{z},t)\,dt + B(\mathbf{z},t)\,d\mathbf{W}_t,
\end{equation}
where $\mathbf{A}$ is the vector of deterministic drift coefficients, $B$ is the matrix of noise amplitudes, and $d\mathbf{W}_t$ is the vector of increments of independent Wiener processes with $\langle d\mathbf{W}_{k,t} \rangle =0$ and $\langle d\mathbf{W}_{k,t}\,d\mathbf{W}_{l,t} \rangle = \delta_{kl}\,dt$.  
The corresponding probability density $\mathcal{P}=\mathcal{P}(\mathbf{z},t)$ obeys the generalized Fokker--Planck equation 
\begin{equation} \label{eq:FP-general-form-multi}
    \frac{\partial \mathcal{P}}{\partial t} = - \sum_i \frac{\partial}{\partial z_i} \left[ A_i(\mathbf{z},t) \mathcal{P} \right] + \sum_{i,j} \frac{\partial^2}{\partial z_i \partial z_j} \left[ \mathsf{D}_{ij}(\mathbf{z},t) \mathcal{P} \right], 
\end{equation} 
where $\mathsf{D} = \tfrac{1}{2}BB^\top$ is the diffusion matrix. 
For a single variable, the diffusion matrix reduces to the diffusion coefficient $D = \tfrac{1}{2}B^2$ in Eq.~\eqref{eq:FP-general-form}. 

\paragraph{The underdamped Langevin equation.} 
The canonical example of a random process is (driven) Brownian motion, e.g., of colloidal particles (like those released from pollen grains~\cite{Brown1828, Einstein1905, Perrin1909}) or single molecules suspended in a solvent, which may furthermore be mechanically stretched~\cite{Liphardt2002, Collin2005Nature}.  
The Langevin equation is an equation of motion for a single Brownian particle in which the surrounding medium exerts a frictional force and a stochastic force modelling random (e.g., thermal) kicks from the environment, 
\begin{equation} \label{eq:Langevin-underdamped}
    m\ddot{x} = - \gamma m\dot{x}  - \partial_x V(x,\lambda) {+f(x,\lambda)} + \xi(t), 
\end{equation}
where the dot denotes a derivative with respect to time, 
$m$ is the mass, $\gamma$ is the damping rate (so that $m\gamma$ is the friction coefficient), $-\partial_x V(x,\lambda)$ is a \textit{conservative} force that arises due to the potential $V(x,\lambda)$, and $f(x,\lambda)$ is a \textit{non-conservative} force applied directly to the particle (that cannot be written as the gradient of a potential, and hence in one dimension is possible for dynamics on a ring with $\oint dx\,f(x,\lambda)\neq0$). 
Both conservative and non-conservative force contributions may depend on time through an external control parameter $\lambda=\lambda(t)$. 

The stochastic force $\xi(t)$ describes randomness due to thermal fluctuations of the environment.  
In the simplest case, the thermal noise is assumed to be Gaussian white noise (that is furthermore independent of $\lambda(t)$ and $f$), characterized by
\begin{equation} \label{eq:Gaussian-white-noise}
    \langle \xi(t) \rangle  = 0 \quad \mbox{and} \quad 
    \langle \xi(t) \xi(t') \rangle  = \frac{2m\gamma}{\beta} \delta(t-t'), 
\end{equation}
where the noise strength is fixed by the \textit{fluctuation-dissipation relation}, which ties the friction coefficient $m\gamma$ (dissipation) to the strength of the noise (fluctuations). 
The Einstein--Smoluchowski relation $D 
= (m\gamma\beta)^{-1}$~\cite{Einstein1905,Smoluchowski1906} recasts the autocorrelation function of the noise as\footnote{Sometimes (e.g., as in Ref.~\cite{Deffner2019book}) the noise autocorrelation function is expressed as $\langle \xi(t)\xi(t')\rangle =2D\delta(t-t')$, where $D$ is the noise strength (not the diffusion coefficient). 
}
\begin{equation}
    \langle \xi(t) \xi(t') \rangle  = 2(m\gamma)^2 D \delta(t-t'). 
\end{equation}

\paragraph{The Klein--Kramers equation.} 
The corresponding Fokker--Planck equation describing the joint probability density for the underdamped Langevin equation is the \textit{Klein--Kramers} equation.  
Comparing the underdamped Langevin equation~\eqref{eq:Langevin-underdamped} with the general multivariate form~\eqref{eq:Langevin-general-form-multi}, where $\mathbf{z}=(x,v)$, 
\begin{equation}
    dx = v\,dt, \qquad dv = \left(\frac{-\partial_x V(x,\lambda) + f(x,\lambda)}{m}- \gamma v\right)dt + \frac{\xi(t)}{m}\,dt, 
\end{equation}
we see that by expressing the noise as $\xi(t)\,dt = \sqrt{2m\gamma /\beta }\,dW_t$, the dynamics correspond to a single Wiener process on the velocity.  
Substituting the drift and diffusion coefficients into the Fokker--Planck equation~\eqref{eq:FP-general-form-multi} yields the Klein--Kramers equation for the probability density $\mathcal{P}=\mathcal{P}(x,v,t)$,
\begin{equation} \label{eq:Klein-Kramers-expanded}
    \frac{\partial}{\partial t}\mathcal{P} = -v\frac{\partial \mathcal{P}}{\partial x} +
    \left(\frac{\partial_x V(x,\lambda)-f(x,\lambda)}{m}\right)\frac{\partial \mathcal{P}}{\partial v} +\gamma  \frac{\partial}{\partial v} (v\mathcal{P}) + \frac{\gamma}{m\beta} \frac{\partial^2 \mathcal{P}}{\partial v^2},  
\end{equation}
whose stationary solution for a time-independent potential and in the absence of a non-conservative force (i.e., for constant $\lambda$ and $f=0$) is the Boltzmann--Gibbs distribution $\propto e^{-\beta\left[mv^2/2 + V(x,\lambda)\right]}$.
The first two terms in Eq.~\eqref{eq:Klein-Kramers-expanded} describe deterministic dynamics, and for $f=0$ reduce to the Liouville equation for Hamiltonian flow of a closed system, while the last two terms describe the friction and thermal diffusion dynamics. 

For a free particle ($V(x,\lambda)=\text{constant},\,f(x,\lambda)=0$), integrating Eq.~\eqref{eq:Klein-Kramers-expanded} over $x$ yields the Fokker--Planck equation for Ornstein--Uhlenbeck dynamics~\cite{UhlenbeckOrnstein1930, risken1989fokker}, with $\mathcal{P}(v,t)= \int dx\,\mathcal{P}(x,v,t)$,
\begin{equation}
    \frac{\partial \mathcal{P}}{\partial t} = \gamma\frac{\partial}{\partial v} (v\mathcal{P}) +\frac{\gamma}{m\beta} \frac{\partial^2 \mathcal{P}}{\partial v^2}, 
\end{equation}
whose stationary solution is the Maxwell--Boltzmann velocity distribution $\propto e^{-\beta mv^2/2}$, i.e., the velocity marginal of the Boltzmann--Gibbs distribution. 

\paragraph{The overdamped Langevin (Smoluchowski) equation.}

The Klein--Kramers equation holds for arbitrary damping rates $\gamma$. 
In the heavily underdamped limit, the damping rate $\gamma$ is small when compared to other dynamical rates, so that for constant $\lambda$ and $f=0$ the dynamics tend to become Hamiltonian as $\gamma\to0$. 
By contrast, in the overdamped limit, $\gamma$ is assumed to be much greater than other dynamical rates, so that the velocity equilibrates almost instantaneously and we can describe dynamics in terms of the position alone,  neglecting $m\ddot{x}$, 
\begin{equation}
    \gamma m \dot{x} = - \partial_x V(x,\lambda) +f(x,\lambda) + \xi(t). 
\end{equation}
As above, by expressing the noise as $\xi(t)\,dt = \sqrt{2m\gamma/\beta}\,dW_t$, the dynamics here correspond to a single Wiener process on the position. 
In the overdamped limit, to leading order, the corresponding Fokker-Planck equation for the position distribution $\mathcal{P}(x,t)= \int dv\,\mathcal{P}(x,v,t)$ is the Smoluchowski diffusion equation, 
\begin{equation}
    \frac{\partial \mathcal{P}}{\partial t} = \frac{\partial}{\partial x}\left[\frac{\partial_x V(x,\lambda) - f(x,\lambda)}{m\gamma} \mathcal{P} + D\frac{\partial \mathcal{P}}{\partial x}\right].  
\end{equation}
For constant $\lambda$ and $f=0$, the stationary solution is the Boltzmann distribution $\propto e^{-\beta V(x,\lambda)}$, i.e., the position marginal of the Boltzmann--Gibbs distribution.

\subsection{Stochastic energetics and entropy}

\paragraph{Stochastic energetics. } Sekimoto's key insight~\cite{Sekimoto1997, Sekimoto1998, Sekimoto2010book} was to apply the first law to an individual fluctuating trajectory. 
{In this section, we restrict ourselves to the overdamped limit (for which the internal energy is specified by the potential at the level of individual trajectories) and adopt the mid-point (Stratonovich) convention (for which the rules of ordinary calculus hold) when evaluating stochastic integrals over the increment $dx$~\cite{Seifert2012}.}

For a particle at equilibrium (i.e., for constant $\lambda$ and $f=0$), no work is applied to the system, so that any change in the internal energy $de$ for a single trajectory $x$ corresponding to the position of the particle in the potential is due to a transfer of heat $\dbar q$ from or to the environment, 
\begin{equation}
    de = dV(x,\lambda) = (\partial_x V(x,\lambda)) dx = \dbar q.
\end{equation}
In order to do work on the particle, we require a time-dependent potential $V(x,\lambda)$ and/or an external force $f(x,\lambda)$, where the explicit time-dependence enters through $\lambda=\lambda(t)$, so that  
\begin{equation} \label{eq:stochastic-work}
    \dbar w = (\partial_\lambda V(x,\lambda)) d\lambda + f\, dx. 
\end{equation}
The first term is the partial derivative of the potential with respect to the externally controllable work parameter $\lambda$, and the second is the work done by the non-conservative force $f$ over the displacement $dx$. 
In the overdamped limit, the change in internal energy is given in terms of the potential, and has contributions from the motion of the particle and the changing potential, $de = dV(x,\lambda) = \partial_x V(x,\lambda)\,dx+\partial_\lambda V(x,\lambda)\,d\lambda$, so that by the first law, the heat absorbed by the system from the surroundings is equivalent to minus the mechanical force applied times displacement
\begin{equation} \label{eq:stochastic-heat}
    \dbar q = dV(x,\lambda) - \dbar w = -(-\partial_x V(x,\lambda) + f)dx. 
\end{equation}

In fact, the stochastic work~\eqref{eq:stochastic-work} falls out of the overdamped Langevin equation
\begin{equation}
    0=-(-m\gamma\dot{x}+\xi_t)dx+\left(\partial_x V(x,\lambda)-f(x,\lambda)\right)dx, 
\end{equation}
where we have separated contributions from interactions with the environment and mechanical forces. 
Simply identifying the heat transferred to the particle with the environmental contributions (i.e., the external terms), $\dbar q = (-m\gamma \dot{x}+\xi_t) dx$, the first law $de = \dbar w + \dbar q$ uniquely defines the stochastic work~\eqref{eq:stochastic-work} for a single trajectory. 

Furthermore, by integrating the stochastic work~\eqref{eq:stochastic-work} and stochastic heat~\eqref{eq:stochastic-heat} along a trajectory over a duration $t$, the work and heat become functionals of the trajectory
\begin{align}
    w[x(\tau)] &= \int_0^t ds\, \left[ \partial_\lambda V(x(\tau),\lambda(\tau)) \dot{\lambda}(\tau) + f(x(\tau),\lambda(\tau)) \dot{x}(\tau)\right] \\
    q[x(\tau)] &= \int_0^t ds\, \left[ \partial_x V(x(\tau),\lambda(\tau)) - f(x(\tau),\lambda(\tau))\right] \dot{x}(\tau),  
\end{align}
where the first law holds on the level of an individual trajectory, 
\begin{equation}
    \Delta e = V(x(t),\lambda(t)) - V(x(0),\lambda(0)) = w[x(\tau)] + q[x(\tau)].
\end{equation}
Note that the heat and work are functionals of the trajectory (cf. inexact differentials).

\paragraph{Stochastic entropy} Can we analogously define the entropy along an individual trajectory? 
For the entropy, there are two contributions. 
First, there is the entropy due to heat transfer from the system to the environment, 
\begin{equation}
    \Delta s^\mathrm{environment}[x(\tau)] \equiv q[x(\tau)]/T, 
\end{equation}
and second, there is the stochastic/trajectory-dependent entropy~\cite{Seifert2005PRL}, 
\begin{equation}
    s(\tau) \equiv -\ln{p(x(\tau),\tau)}, 
\end{equation}
which ultimately depends on the ensemble since we need to solve the Fokker--Planck equation to obtain the probability $p(x,\tau)$, and evaluate along the trajectory $x(\tau)$.

The total entropy production along a trajectory is given by 
\begin{equation}
    \Delta s^\mathrm{total} = \Delta s^\mathrm{environment}+\Delta s, 
\end{equation}
where
\begin{equation}
    \Delta s \equiv -\ln{p(x_t,\lambda_t)} + \ln{p(x_0,\lambda_0)} 
\end{equation}
obeys the integral fluctuation theorem~\eqref{eq:integral-fluctuation-theorem}, 
\begin{equation}
    \langle \exp{(-\Delta s^\mathrm{total})} \rangle = 1, 
\end{equation}
for arbitrary initial distribution, time-dependent driving, and process duration~\cite{Seifert2005PRL,Seifert2012}. 

\subsection{Crooks relation}
Another very important fluctuation theorem is the Crooks relation~\cite{Crooks1998, Crooks1999PRE}, which relies on the principle of detailed balance in systems coupled to a thermal reservoir. 

Crooks fluctuation theorem relates the probability densities of the work performed on the system in different realizations in the forward and reverse driving processes, $\mathcal{P}_F(+W)$ and $\mathcal{P}_R(-W)$, as
\begin{equation} \label{eq:Crooks-relation}
    \mathcal{P}_R(-W) = \exp{(-\beta(W-\Delta F))} \mathcal{P}_F(+W), 
\end{equation}
and holds arbitrarily far from equilibrium for microscopically reversible dynamics. 
Both forward and reverse processes begin at thermal equilibrium set by the inverse temperature $\beta$, at $\lambda_0$ and $\lambda_\tau$, respectively, where $\tau$ is the total duration of the driving protocol, and the reversed protocol is specified by $\lambda_R(t) = \lambda_F(\tau-t)$. 

The Crooks relation implies the Jarzynski equality, and has been verified experimentally in a variety of settings, spanning from RNA folding and unfolding experiments~\cite{Collin2005Nature} to single-electron devices~\cite{Saira2012}. 

\bibliographystyle{fqxi_foundational_review}
\bibliography{notions_qmwork}
\end{document}